\documentclass[10pt]{wlscirep}
\usepackage[utf8]{inputenc}
\usepackage[T1]{fontenc}
\usepackage{adjustbox}
\usepackage{subfigure}
\usepackage{color, colortbl}
\usepackage{soul}

\title{Analyzing the Impact of the Automatic Ball Strike System in Professional Baseball through a Case Study on KBO League Data}

\author[1]{Kichang Lee}
\author[2]{Kyungsik Han}
\author[1,3,*]{JeongGil Ko}
\affil[1]{Yonsei University, School of Integrated Technology, Seoul, 03789, Republic of Korea}
\affil[2]{Hanyang University, Department of Data Science, Seoul, 08031, Republic of Korea}
\affil[3]{POSTECH, Graduate School of Artificial Intelligence, Pohang, 37673, Republic of Korea}
\affil[*]{jeonggil.ko@yonsei.ac.kr}

\newcommand{\rev}[1]{{\textcolor{black}{#1}}}

\begin{abstract}

Recent advancements in professional baseball have led to the introduction of the Automated Ball-Strike (ABS) system, or ``robot umpires,'' which utilize machine learning, computer vision, and precise tracking technologies to automate ball-strike calls. The Korean Baseball Organization (KBO) league became the first professional baseball league to implement ABS during the 2024 season.
\rev{Leveraging pitch data from 2,515 KBO games across multiple seasons and employing mathematical modeling, we examine the aggregate decision tendencies of human umpires versus those of the ABS within the ``gray zone'' of the strike zone.}
We propose and answer four research questions to examine the differences between human and robot umpires, player adaptation to ABS, assess the ABS system's fairness and consistency, and analyze its strategic implications for the game. Our findings offer valuable insights into the impact of technological integration in sports officiating, providing lessons relevant to future implementations in professional baseball and beyond.

\end{abstract}
\begin{document}

\flushbottom
\maketitle

%
%

\section*{Introduction}
\label{sec:intro}

The accuracy and consistency of umpire decisions in professional baseball are crucial for the game's integrity. Recent technological advancements have introduced methods to assist or replace human judgment, such as the Automated Ball-Strike (ABS) system, commonly known as ``robot umpires.'' The ABS system uses machine learning~\cite{sugizaki2023umpire, thomas2017computer}, computer vision~\cite{chen2010contour, lee2020method}, and precise tracking technologies~\cite{hsieh2024neural, lage2016statcast} to determine balls and strikes, potentially offering more consistency and less bias than human umpires. The Korea Baseball Organization (KBO) league was the first professional league to adopt ABS technology, and this work examines its impact through an analysis of pitch data.


The ABS system employs cameras, radars, and software to track a pitched baseball's trajectory, speed, and position relative to the strike zone~\cite{gueziec2002tracking, tseng2024pitching}. The \textit{strike zone}, defined by the batter's stance and league rules, is a three-dimensional space that the ABS can measure in real-time to eliminate the subjectivity of human umpiring.

Traditionally, umpires rely on their judgment, experience, and training to call balls and strikes. Despite rigorous training, human error leads to occasional disputes and inconsistencies~\cite{russell1997concept, kim2014seeing, buUmpiresMissed}. \rev{Williams~\cite{buUmpiresMissed} reported that Major League Baseball umpires made 34,294 incorrect ball–strike calls ($\approx$14 per game) during the 2018 season. He further attributed these errors to various human biases, including game context, umpire age, and level of experience.} Players sometimes exploit these inaccuracies by testing the umpire's tendencies early in the game~\cite{macmahon2008contextual} or catchers ``framing'' pitches to make them appear as strikes~\cite{sullivan2015astros, deshpande2017hierarchical, judge2018bayesian}. ABS aims to mitigate such tactics by bringing objectivity and consistency to ball-strike calls, reducing human bias. Supporters argue that ABS ensures a fairer game, eliminating subjective errors and providing precise, consistent decisions. Additionally, ABS can enhance baseball analytics, offering insights into player performance and strategies.

The KBO League's integration of the ABS system in 2024 reflects a significant shift toward modernizing the game and enhancing its fairness and accuracy~\cite{nbcnewsTechnicalDifficulties, espnWhenWill}. This decision was driven by the desire to reduce umpiring errors, improve consistency, and enhance the fan experience~\cite{ynaExpandsStrike}. The league conducted extensive trials and gathered feedback from players, coaches, and umpires to ensure a smooth transition~\cite{dongaKoreanBaseball}. Implementing the ABS system positions the KBO League at the forefront of technological adoption in professional baseball, marking a pivotal change in the history of the game. Following the KBO's lead, Major League Baseball (MLB) began using the ABS-based challenge system in Triple-A games starting June 25, 2024~\cite{nytimesTripleAGames}.


\rev{However, the shift to ABS also introduces important adaptation challenges. Specifically, experienced players in the KBO reported frustration with the tighter, more uniform strike zone, necessitating changes in pitching and batting strategies~\cite{shorturlPreparesRobot}. For pitchers, the precise and consistent zone may lead to more predictable outcomes and influence pitch selection, while batters can adjust their tactics at the plate knowing that the strike zone will be enforced without human variability. Fans also faced a learning curve in tracking the new boundaries before coming to appreciate the system’s objectivity~\cite{shorturlUnpacksWhat}. Nevertheless, before understanding these impacts on player performance and fan experience, it is essential to first analyze the fundamental differences between ABS and human umpiring. Accordingly, our study focuses on quantifying and characterizing divergences between ABS and human strike zones, thereby laying the groundwork for future investigations into how these technological changes affect gameplay and spectator engagement.}


\rev{This research analyzes pitch data from three and a half KBO League seasons (three pre-ABS seasons and the first half of the ABS-adopted 2024 season) to compare decision patterns of human and robot umpires, with particular emphasis on the so-called ``gray zone,'' denoting the area in which the human umpire’s estimated strike zone and the ABS strike zone differ most significantly. By quantifying these boundary mismatches, our analysis aims to pinpoint where adaptation will be potentially most critical for players and teams.}
Specifically, we address four research questions:


\noindent{$\bullet$ \textbf{ RQ1:} Is there a significant difference between the calls made by human umpires and robot umpires?}

\noindent{$\bullet$ \textbf{ RQ2:} Where does the gray zone in strike-ball calls lie, and how does the strike call differ for robot umpires?}

\noindent{$\bullet$ \textbf{ RQ3:} Does the data show how players adjust to ABS?}

\noindent{$\bullet$ \textbf{ RQ4:} Is ABS really fair and consistent as expected?}

\rev{To the best of our knowledge, previous strike-zone analyses have focused primarily on biases in human umpiring or player behavior~\cite{flannagan2024psychophysics, whiteside2016ball}, whereas our study uniquely examines the robot umpire's call patterns.}
The findings from this research will contribute to the discussion on integrating technology in sports officiating. Insights from the KBO's experience with ABS will provide valuable lessons for other leagues considering similar technologies.
\rev{Specifically, this study aims to refine our understanding of how emerging officiating technologies might balance tradition and innovation in professional baseball. Through an examination of the ABS system’s strike-zone characteristics, we offer a preliminary assessment of potential implications and establish a foundation for future research into technology’s role in the game.}


\section*{Methodology}
\label{sec:method}

In this section, we discuss the detailed specifics of the data preparation process and our methodology for analysis.

\subsection*{Data Collection}

To compare strike-ball calls made by human and robot umpires, we compiled data from Naver Sports~\cite{naver} covering 2,515 games over four years (2021 to 2024), totaling 22,934 innings. This dataset includes 941 players from 10 teams and 42 home plate umpires. While each of the ten KBO teams plays 144 games per season, discrepancies in the database led to a few missing games. We analyzed data from the 2024 season with the ABS system up to the All-Star break, and compared it with human umpire data from 2021 to 2023. 
Notably, the strike zone was officially expanded in 2022, resulting in a narrower zone in 2021 compared to later years. \rev{While this change precludes direct one‐to‐one comparisons of absolute boundaries, we nonetheless include the 2021 human-umpire data to capture and characterize inherent bias and decision-making patterns that persist across seasons. We then evaluate the impact of ABS adoption by comparing data from 2022–2023 and 2024.}
\rev{Additionally, we included the 2024 season Major League Baseball (MLB) pitch data, sourced from Statcast Savant~\cite{mlbBaseballSavant}, solely for evaluating cross-league generalization as described in the \textit{Discussion} section. Note that unless explicitly mentioned as the MLB data, all references to 2024 data in this paper refer to robot umpire calls in the KBO League.}



\begin{table}[!ht]
\begin{adjustbox}{width=0.9\linewidth,center}
\centering
\begin{tabular}{|c|c|c|c|cc|}
\hline
\hline
\textbf{Year} & \textbf{2021}                    & \textbf{2022}  & \textbf{2023}  & \multicolumn{2}{c|}{\textbf{2024}}                      \\ \hline
\# Games      & 701                              & 704            & 702            & \multicolumn{1}{c|}{408}            & 1,016             \\ \hline
\# Innings    & 6,330                            & 6,447          & 6,530          & \multicolumn{1}{c|}{3,727}          & 9,273             \\ \hline
\# Pitches    & 207,581                          & 204,084        & 203,666        & \multicolumn{1}{c|}{122,632}        & 299,421           \\ \hline
\# Pitchers   & 305                              & 279            & 284            & \multicolumn{1}{c|}{251}            & 658               \\ \hline
\# Batters    & 281                              & 274            & 290            & \multicolumn{1}{c|}{254}            & 541               \\ \hline
Duration      & 4/3$\sim$7/11, 8/10$\sim$10/31   & 4/2$\sim$10/11 & 4/1$\sim$10/17 & \multicolumn{1}{c|}{3/23$\sim$7/4} & 4/1$\sim$6/18     \\ \hline
Remarks       & KBO, human umpire, narrow strike zone & KBO, human umpire   & KBO, human umpire   & \multicolumn{1}{c|}{KBO, ABS}            & MLB, human umpire \\ \hline
\hline
\end{tabular}
\end{adjustbox}
\caption{Detailed dataset configuration.} 
\label{tab:dataset_config}
\end{table}

\rev{Table~\ref{tab:dataset_config} provides an overview of our dataset, including the number of games and total pitches in each season, as well as counts of unique pitchers, batters, and umpiring systems (human vs. ABS). While Table~\ref{tab:dataset_config} summarizes these high-level statistics, each individual pitch record in our data also contains detailed attributes, pitch speed, pitch type, location relative to the strike zone, pitch outcome (e.g., ball, strike, foul, hit), and context meta data (player, umpire, inning, ball count, etc), which are used for our analyses. Throughout this paper, the terms `robot umpire' and `ABS' are used interchangeably.} The dataset will be publicly accessible at \href{https://github.com/eis-lab/where-do-the-robot-umpires-see}{https://github.com/eis-lab/where-do-the-robot-umpires-see}.

\begin{figure}[!ht]
    \centering
    \includegraphics[width=0.75\linewidth]{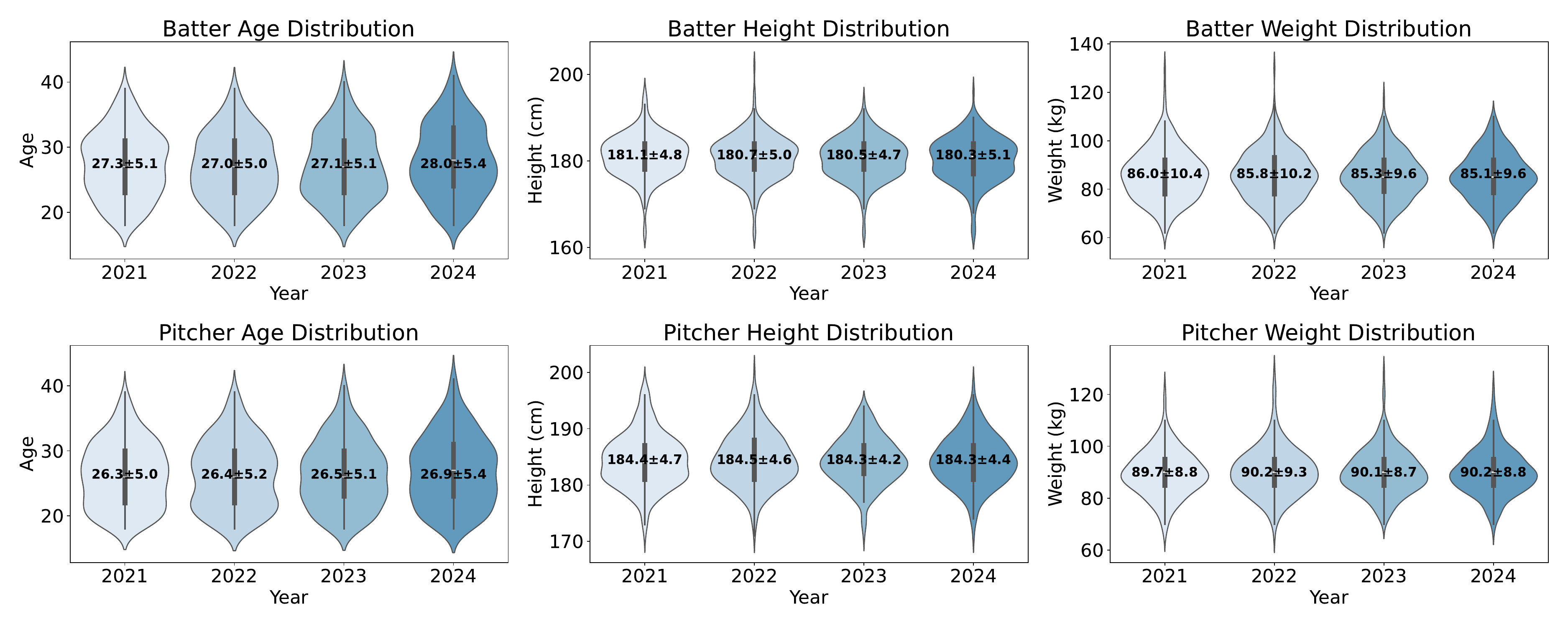}
    \caption{Age, height, and weight distribution of KBO players in different years}
    \label{fig:player_stats}
\end{figure}

Figure~\ref{fig:player_stats} shows the distribution of pitchers' and batters' ages, heights, and weights in the KBO datasets. The plots reveal no significant statistical differences across years, indicating that the analysis is not notably influenced or biased by variations in players' physical characteristics.



\subsection*{Method}

We analyze the pitch data by examining statistical differences across the years. First, we address \textbf{RQ1}: ``Is there a significant difference between the calls made by human umpires and robot umpires?'' by modeling the strike zone using a parametric logistic regression model and perform a statistical analysis of its parameters. The details of this strike zone model and results are presented in \textit{Analysis I}.

Second, to address \textbf{RQ2}: ``Where does the gray zone lie and what is the difference?'', we analyze the gray zones identified from RQ1, considering pitch locations, types, and batter stances in \textit{Analysis II}. We quantify and examine the ratio of strike and ball calls for pitches thrown in these gray zones.

In \textit{Analysis III}, we explore \textbf{RQ3}: ``Are the players adjusting to ABS?'' by investigating changes in players' strategies and reactions. This analysis assesses whether players are adapting to robot umpires.

Finally, in \textit{Analysis IV}, we address \textbf{RQ4}: ``Is ABS really fair and consistent as expected?'' We conduct an in-depth analysis of umpire calls to evaluate the fairness and consistency of human versus robot umpires.

\subsection*{Strike Zone Modeling}
\label{subsec:zone-modeling}
\begin{figure}[!t]
    \centering
    \includegraphics[width=0.65\linewidth]{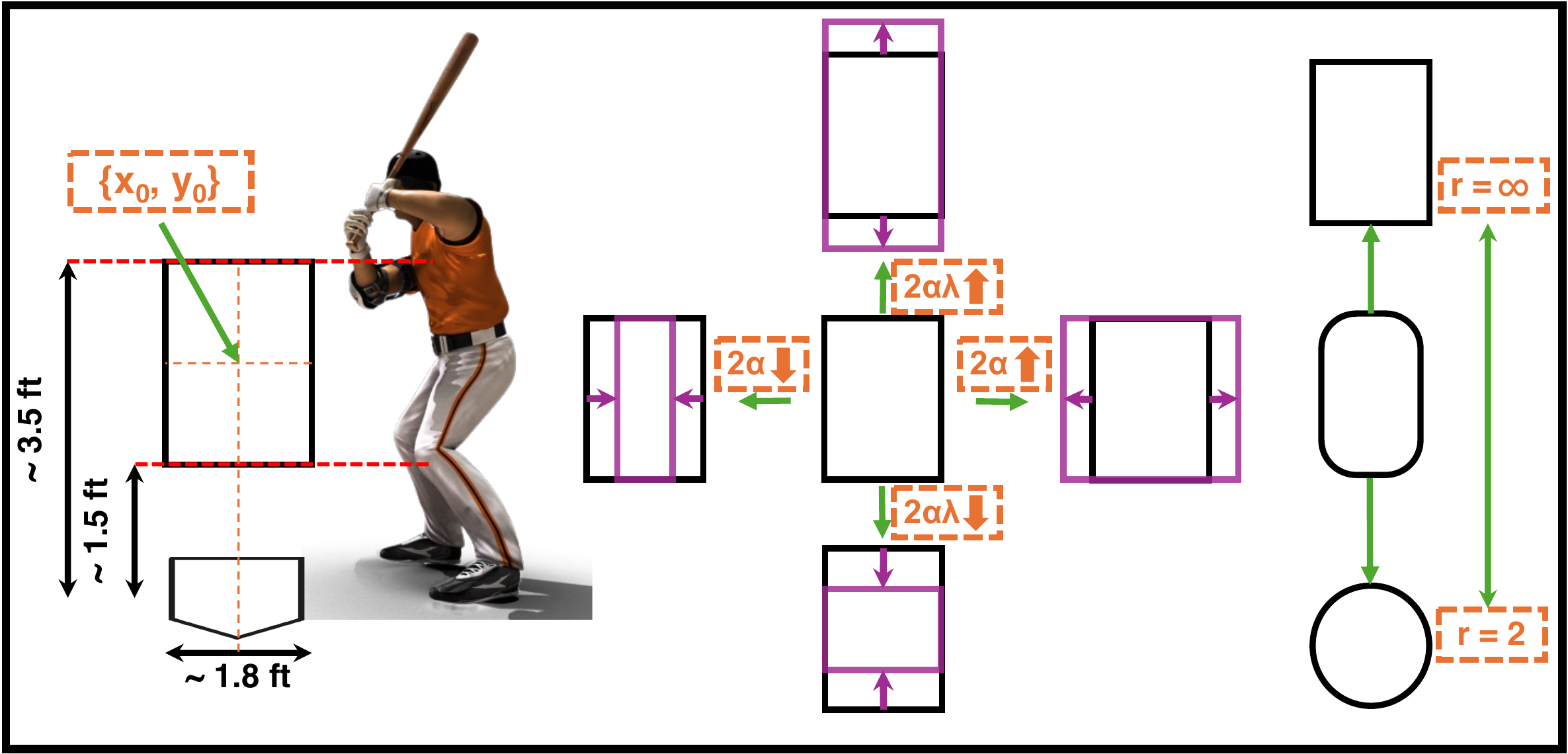}
    \caption{Illustration of relationship between parameters adapted in the parametric model. $\{x_{0},y_{0}\}$ denotes the center of the strike zone, $2\alpha$, $2\alpha \lambda$ indicates the width and height, respectively. and $r$ controls the rectilinearity of the strike zone.}
    \label{fig:strikezone}
\end{figure}

To model the strike zone, we borrow the parametric logistic regression model proposed by Flannagan et al. ~\cite{flannagan2024psychophysics} as follows.

\begin{equation}
\centering
    P(``Strike")=logit^{-1}(-\beta (d-\alpha))
    \label{eq:pstrike}
\end{equation}

\rev{In Eq~\ref{eq:pstrike}, $P(\text{``Strike"})$ denotes the probability of classifying a pitch as a ``strike'' rather than a ``ball''. $logit^{-1}(\cdot)$ is the logistic function ($=\frac{\exp(x)}{1+\exp(x)}$), which maps the input to a probability.} $\beta$ affects the steepness of the probability contour, determining how the probability transitions between a strike and a ball based on the distance ($d$) from the zone's center. Parameter $\alpha$ defines the boundary size of the zone, influencing its width and height.
%
%
The distance measure $d$ is defined as:

\begin{equation}
\centering
d = \sqrt[r]{|(x-x_{0})|^{r} + |(\frac{y-y_{0}}{\lambda})|^{r}}
\label{eq:d}
\end{equation}

where $x$ and $y$ correspond to the horizontal and vertical location of a pitch measured in feet. 
The origin of the coordinate system is based on the ground level for the vertical axis and the home plate center for the horizontal axis. $x_{0}$ and $y_{0}$ denote the center of the zone, serving as the reference point for measuring \rev{distance $d$}. $\lambda$ is a scaling factor that adjusts the ratio of the vertical and horizontal dimensions of the zone, while $r$ fine-tunes the curvature of the strike zone. 

As noted by Flannagan et al., each parameter can be interpreted to quantify strike zone characteristics~\cite{flannagan2024psychophysics}. Figure~\ref{fig:strikezone} illustrates the relationship between the parameters and the strike zone morphology. By fitting the function to real-world pitch data, we compute the \textit{center of the zone} ($x_{0}, y_{0}$), and the \textit{width} and \textit{height} of the strike zone ($2\alpha$ and $2\alpha\lambda$, respectively). Here, $\alpha$ and $\alpha\lambda$ represent the half-width and half-height of the zone, with $\lambda$ adjusting the width-height ratio. Thus, the full width and height are twice these values.

Moreover, $\beta$ provides insight into the \textit{consistency} of the zone: a larger $\beta$ indicates that pitches far from the center are less likely to be strikes, and pitches at similar distances from the center receive consistent calls. Lastly, $r$ indicates the \textit{rectilinearity} of the strike zone. \rev{We adopt a p-norm framework to flexibly capture the zone’s geometry. The p-norm generalizes Euclidean distance (p=2) toward the Chebyshev (p$\rightarrow$$\infty$) norm, allowing shapes to interpolate continuously between circular and perfectly rectangular contours.} Analogous to the p-norm, a larger $r$ suggests a shape closer to a quadrilateral. Since the KBO's strike zone is rectilinear, a well-modeled zone should exhibit a high $r$.

\rev{We note that our strike‐zone modeling, based solely on two‐dimensional pitch coordinates and observed call labels, is a generalizable approach both for human and robot umpires. We apply this method separately to the human‐umpire data and to the ABS data for each season in our dataset to produce comparable zone estimates.}

\subsection*{Defining the KBO Rule Book Strike Zone}
\begin{equation} 
Call = \begin{cases} 
       Strike & \text{if }x\in(-0.9,0.9) \text{ and } y\in(1.5,3.5)\\ 
       Ball   & \text{otherwise}
 \end{cases}
 \label{eq:rbsz}
\end{equation}


We define the ground truth strike zone for our work as shown in Eq~\ref{eq:rbsz} and discuss how specific values are determined below.

Firstly, the width of the KBO strike zone is approximately 47.18 cm. It is essential to note that a pitch is deemed a strike if any part of the ball passes through the strike zone. Consequently, we must consider the radius of the ball, which is 3.7 cm (the average radius of a KBO baseball), on each side. Thus, the total width of the strike zone, as defined by the rule book, is 54.58 cm (approximately 1.79 ft). For simplicity, we round this to 1.8 ft (approximately 55 cm), resulting in a horizontal range of (-0.9, 0.9) ft, with the horizontal center at 0.

Determining the strike zone height is more complex due to its adaptive nature, which varies with the batter's height and stance. As an approximation, we refer to the KBO rule book and consider the average height of the league's batters. According to the KBO, the top and bottom boundaries of the strike zone are 56.35\% and 27.64\% of the batter's height, respectively~\cite{ynaExpandsStrike}.

Figure~\ref{fig:player_stats} shows that the average height of batters in the KBO is approximately 180 cm. Based on the specified percentages, the top and bottom boundaries of the strike zone are calculated to be 101.43 cm and 49.75 cm, respectively. To account for the baseball's radius, we adjust these vertical boundaries: the bottom is set at 1.5 ft (approximately 45.75 cm) and the top at 3.5 ft (approximately 105.13 cm). ABS evaluates pitches using two planes: one at the middle of the home plate and another at the back of the strike zone, classifying a pitch as a strike only when passing both planes. For simplicity, our analysis focuses solely on the 2D projection of the pitch location at the middle of the plate.


\section*{Analysis I: Identifying The Gray Zone}
\label{sec:analysis1}
\begin{figure}[!t]
    \centering
    \subfigure[Visualization of the contour line of strike zones from the umpire's point of view over various years.]{
    \includegraphics[width=0.25\linewidth]{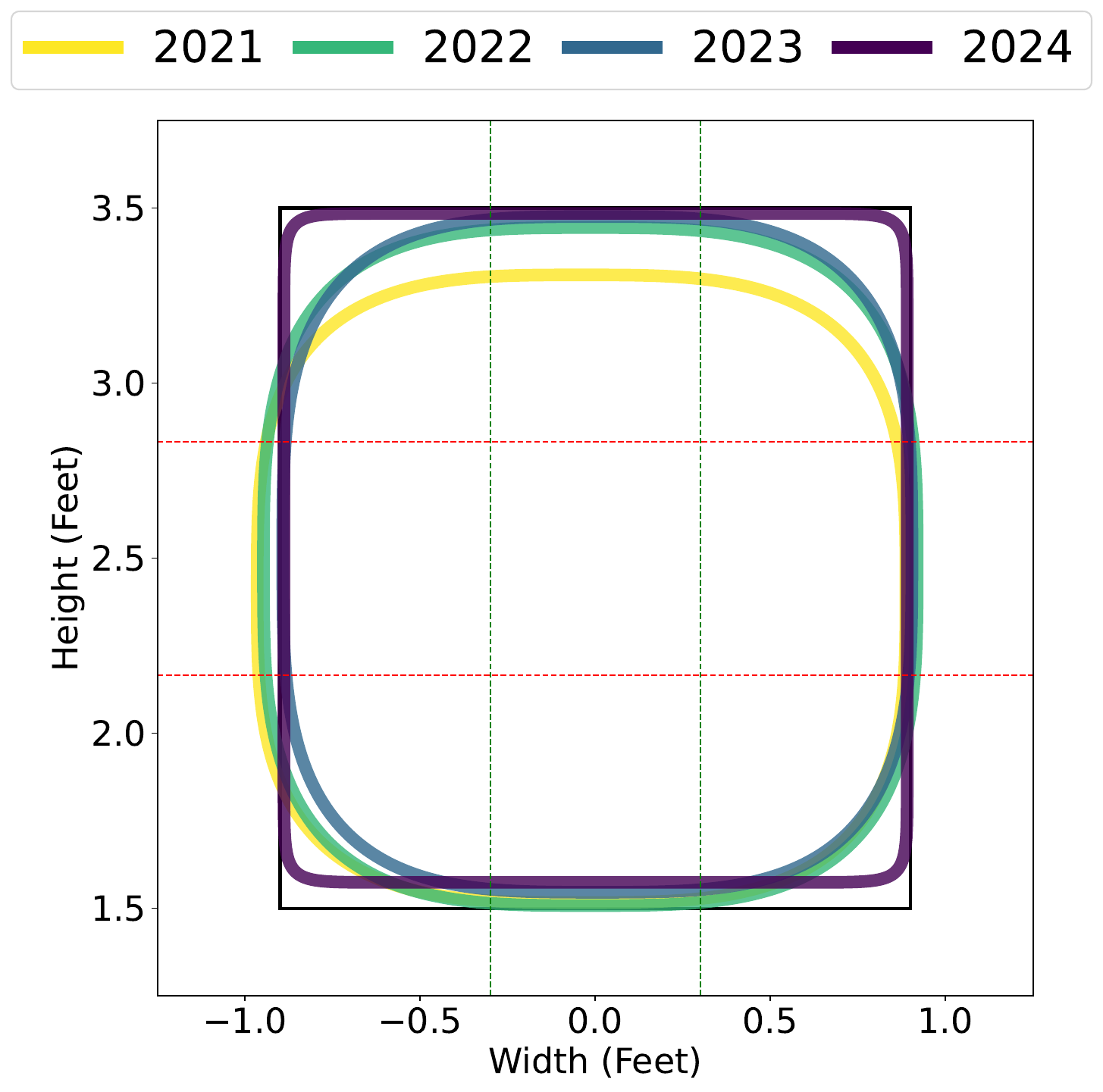}
    }
    \hspace{1ex}
    \subfigure[Distribution of optimized parameters with respect to different years.]{
    \includegraphics[width=0.45\linewidth]{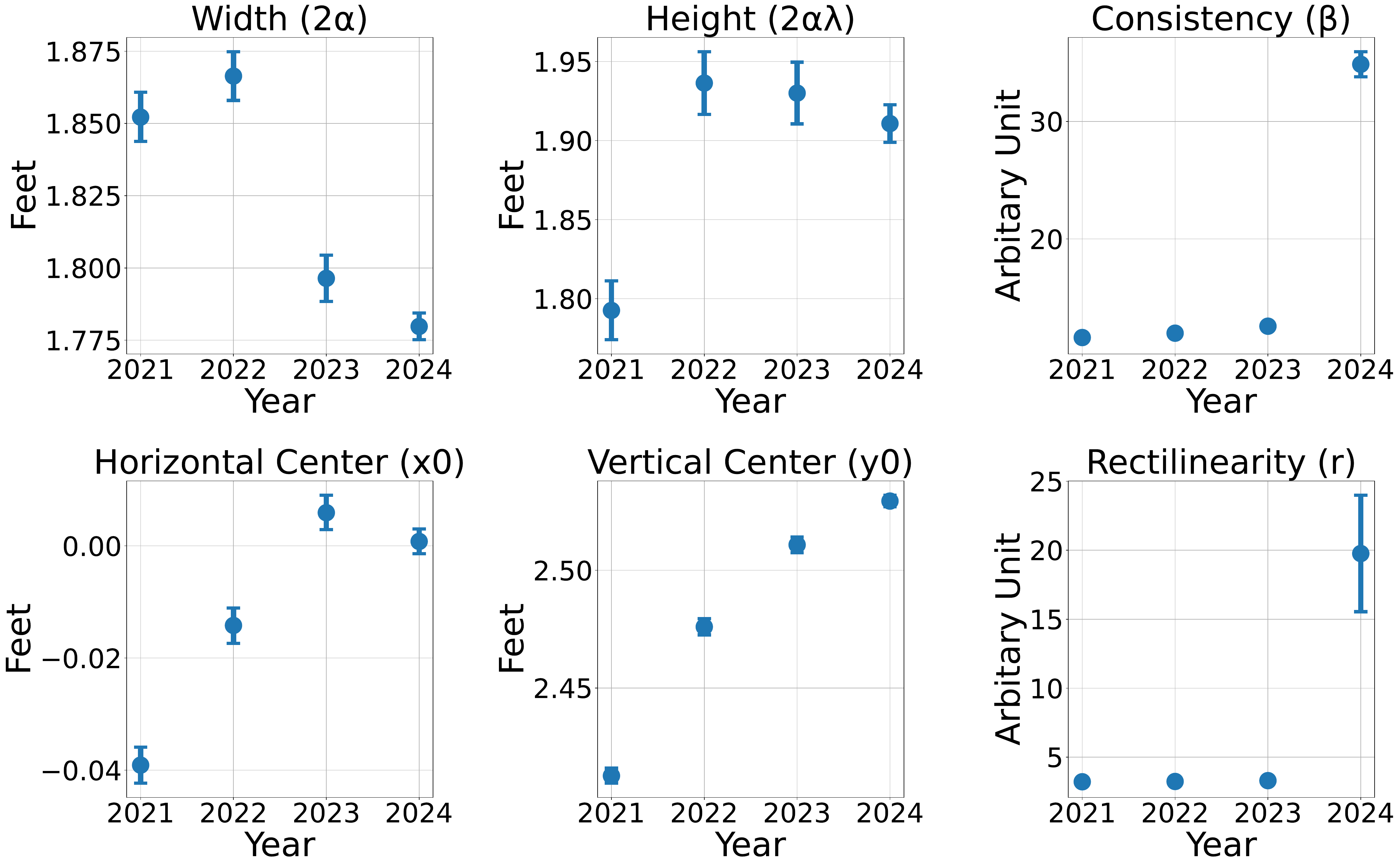}
    }
    \caption{(a) Time trend of the contour of the strike zones. The figure plots the decision boundary (where the probability for calling a pitch as a strike is 0.5) of the fitted model which is identical to the 50\% contour line for each year from 2021 to 2024. The black box in the plot depicts the rule book strike zone. (b) Trend of the fitted model parameter for each year from 2021 to 2024. The vertical bars in the plot illustrate the credible intervals in 95\%}
    \label{fig:human-vs-robot}
\end{figure}

In this section, we target to address our first research question: `\textit{`Is there a significant difference between the calls made by human umpires and robot umpires?''} We qualitatively and quantitatively compare the strike zones of human umpires and the ABS system using pitch data from four different years.

\begin{figure}[!t]
    \centering
    \includegraphics[width=0.98\linewidth]{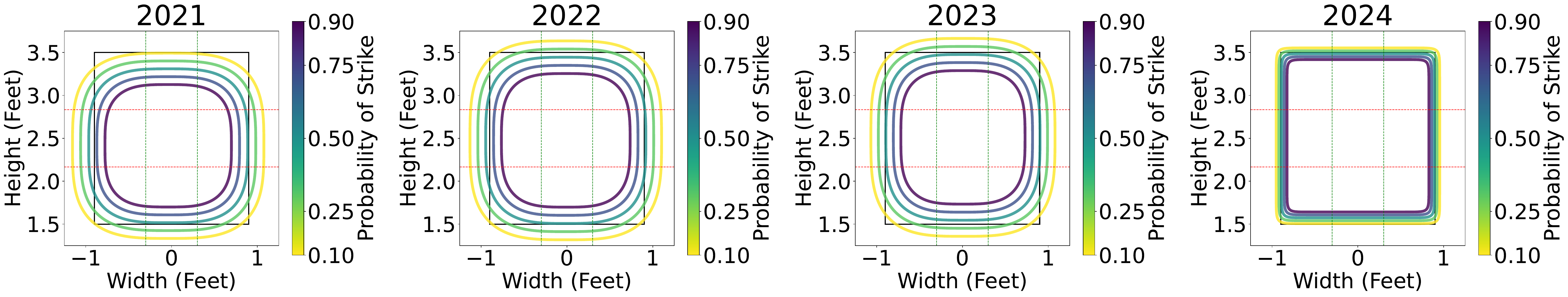}
    \caption{Visualization of the contour lines of strike zones with varying probability value thresholds from the umpire's point of view for each year from 2021 to 2024. A lower probability threshold (\(p \rightarrow 0\)) indicates a tendency to classify the given pitch as a ball, resulting in a contour line that covers a wider area. The black box in the plot represents the rule book strike zone.}
    \label{fig:szcontours}
\end{figure}

Figure~\ref{fig:human-vs-robot}~(a) visualizes the contour lines of strike zones for different years, fitted by the parametric model, while Figure\ref{fig:human-vs-robot}~(b) plots the distribution of each critical parameter from 2021 to 2024. Notably, the strike zone corners reveal definite differences, with the ABS strike zone in 2024 showing a more rectilinear shape compared to the years when human umpires made the calls. This observation is supported by the significantly larger value of $r$ depicted in Figure\ref{fig:human-vs-robot}~(b).

Additionally, Figure~\ref{fig:human-vs-robot}~(a) demonstrates that the strike zones of human umpires (2021-2023) are relatively wider along the horizontal axis compared to ABS. This is quantitatively confirmed by the value of $2\alpha$ in Figure\ref{fig:human-vs-robot}~(b). Regarding the center of the strike zone, ABS exhibits better calibration on the horizontal axis (closer to 0.0) and is positioned slightly higher on the vertical axis, as quantified in Figure\ref{fig:human-vs-robot}~(b).



Figure~\ref{fig:szcontours} presents the contour lines with different strike call probability thresholds. The significantly high $\beta$ in Figure~\ref{fig:human-vs-robot}~(b) confirms that the robot umpire has a relatively dense and strict boundary, indicating that the ABS system possesses not only a clear but also a consistent strike zone. In contrast, the strike zones of human umpires show sparse contour lines, suggesting that their calls were inconsistent and had ambiguous decision boundaries.

We summarize our observations as follows:
(i) There are distinct differences between strike zones determined by human and robot umpires;
(ii) The robot umpire's strike zone is characterized by greater consistency and strictness;
(iii) The strike zone defined by the robot umpire is relatively narrow horizontally, more rectilinear, and shifted toward a comparatively higher location.



\section*{Analysis II: Deeper Understanding of The Gray Zone}
\label{sec:analysis2}

\begin{figure}[t!]
    \centering
    \includegraphics[width=0.7\linewidth]{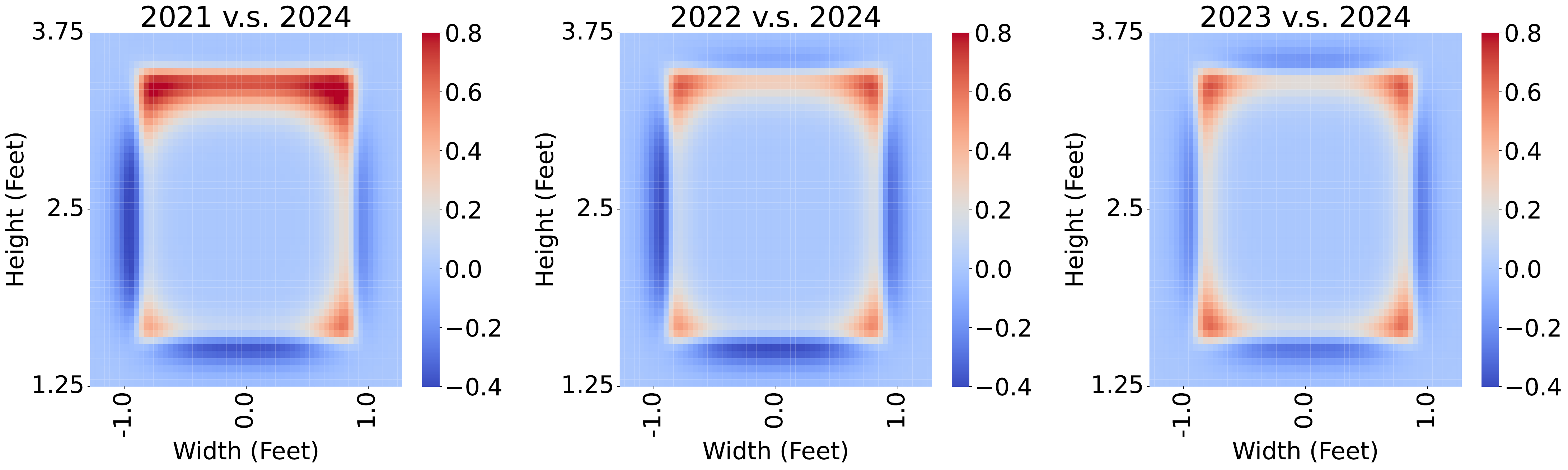}
    \caption{Difference of calculated probability between the human umpires and the robot umpire. The higher/lower value (red/blue) indicates that the robot umpire assigned a high/low probability of strike for the pitches located in the region as strike/ball while the human umpire did not. Best viewed in color.}
    \label{fig:year_diff}
\end{figure}

We now delve into the gray zone observed in \textit{Analysis I} by focusing on a specific zone of interest for an in-depth analysis of the calls made by the ABS system. Figure~\ref{fig:year_diff} visualizes the differences in strike call probabilities between human and robot umpires, highlighting that the major discrepancies occur at the strike zone boundaries. To better analyze these gray zones, we divide the strike zone into four categories: 'high', 'low', 'in', and 'out'. Each gray zone is further subdivided based on whether a given pitch was located within the rule book strike zone or not. Specifically, we define the width of this interest zone as 0.25 feet across all zones.

\begin{figure}[t!]
    \centering
    \includegraphics[width=0.7\linewidth]{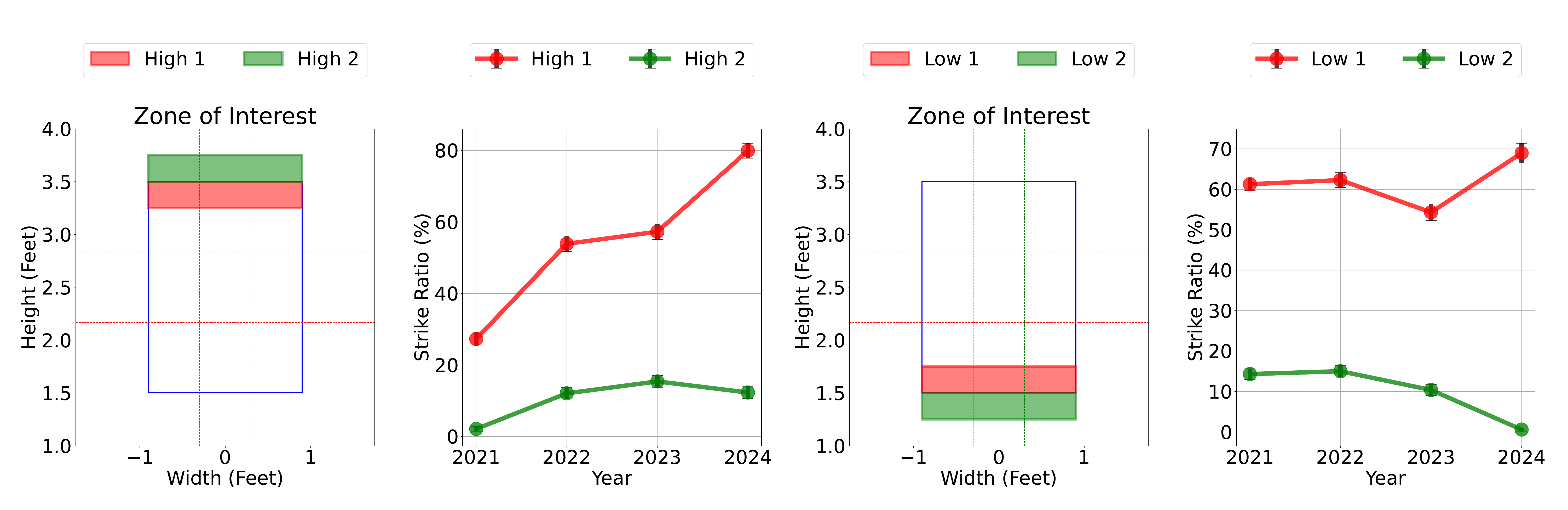}
    \caption{\rev{Visualization of interest zones and trends of strike-ball ratio (\%) over different years with respect to high and low interest zones. The vertical bars in the plot illustrate the 95\% credible intervals. Best viewed in color.}}
    \label{fig:vertical_zone}
\end{figure}

Figure~\ref{fig:vertical_zone} depicts the zones of interest for high and low balls on or near the strike zone. The red-highlighted boxes indicate the zones where the ball is located inside the rule-book strike zone (i.e., High 1 and Low 1), while the green-highlighted boxes indicate zones where the ball is located outside the rule-book strike zone (i.e., High 2 and Low 2). As the second plot in Figure~\ref{fig:vertical_zone} shows, the strike call ratio for pitches located in the high zones within the rule-book strike zone for the 2024 season shows a noticeable increase compared to years when human umpires made the calls (57.28\% in 2023 to 79.77\% in 2024). Additionally, for low balls, the strike rate for pitches that actually entered the strike zone showed a significant increase compared to non-ABS years (62.48\% in 2022 and 54.43\% in 2023, to 68.83\% in 2024, $p<.05$). Interestingly, the number of misjudgments for balls outside the zone decreased (10.49\% in 2023 to 0.58\% in 2024, $p<.05$). While the height of the strike zone is affected by the height of the batters, our statistics still indicate a significant difference between human and robot umpires.


\begin{figure}[t!]
    \centering
    \subfigure[Strike ratios for different pitch types and different years with respect to the pitches located in the high courses.]{
    \centering
    \includegraphics[width=0.97\linewidth]{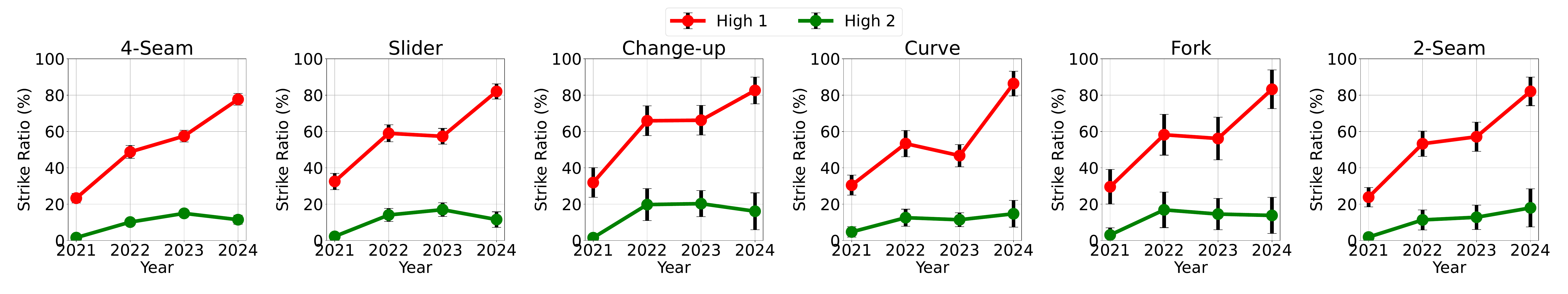}
    }
    \subfigure[Strike ratios for different pitch types and different years with respect to the pitches located in the low courses.]{
    \centering
    \includegraphics[width=0.97\linewidth]{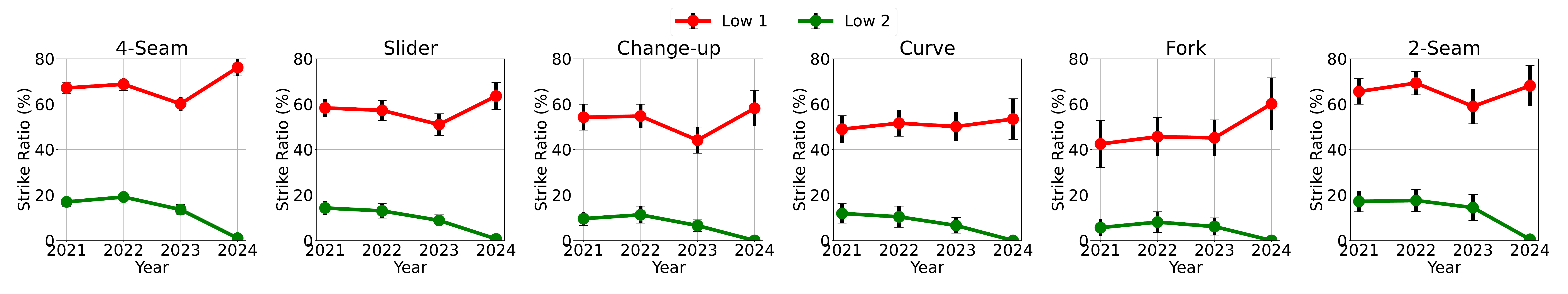}
    }    
    \vspace{-2ex}
    \caption{\rev{Trend of strike call ratios concerning the pitch location for different pitch types. The vertical bars in the plots indicate the 95\% credible interval.}}
    \label{fig:vertical_stuff}
\end{figure}

We further explore the strike call ratio for the top six pitch types: 4-seam fastball, slider, change-up, curveball, forkball, and 2-seam fastball. Figure~\ref{fig:vertical_stuff} shows the strike ratio over four seasons for these pitch types in the zones defined in Figure~\ref{fig:strikezone}. For high-zone pitches in Figure~\ref{fig:vertical_stuff}~(a), all types show a statistically significant increase in strike ratio, with the curveball exhibiting the largest increase (46.46\% to 84.56\%). This suggests human umpires had difficulty classifying high curveballs due to their vertical drop near the plate. For low-zone pitches in Figure~\ref{fig:vertical_stuff}~(b), the 4-seam fastball shows a significant difference in the 'Low 1' zone. In the 'Low 2' zone, all pitch types show a decrease in strike calls, indicating ABS's stricter adherence to the rule book compared to human umpires. The 4-seam and 2-seam fastballs show the most pronounced decrease in strike ratios ($p<.05$): 13.71\% to 1.08\% for the 4-seam and 14.26\% to 0.41\% for the 2-seam from 2023 to 2024.

\begin{figure}[t!]
    \centering
    \includegraphics[width=0.7\linewidth]{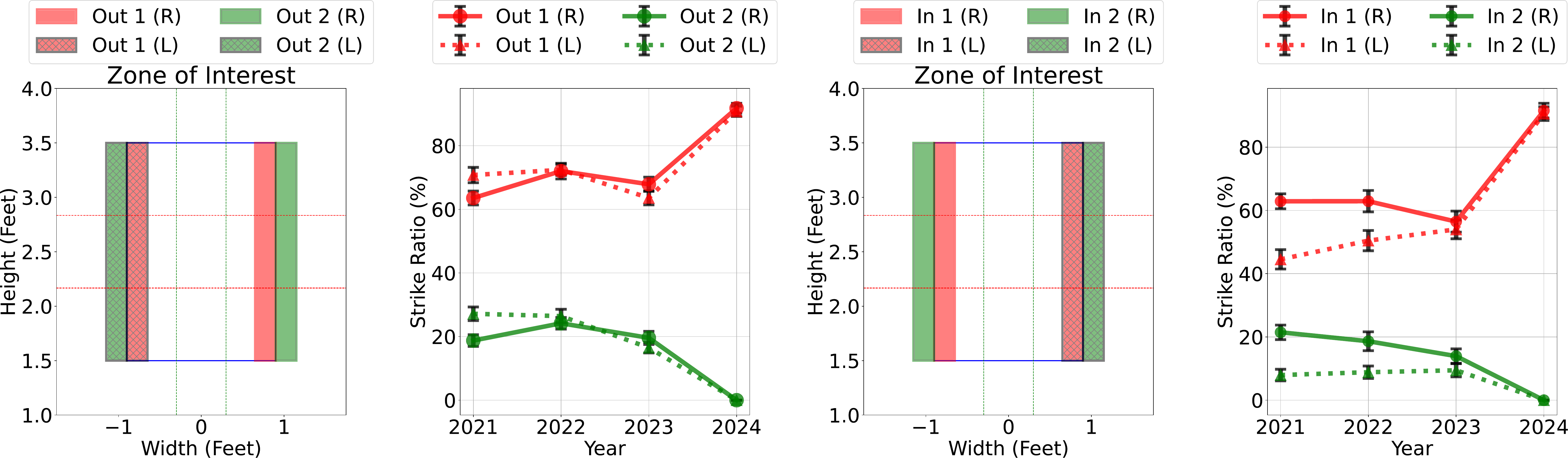}
    \caption{Visualization of interest zones and trends of strike-ball ratio (\%) over different years with respect to the interest zone. The vertical bars in the plot illustrate the 95\% credible intervals. Best viewed in color.}
    \label{fig:horizontal_zone}
\end{figure}

We analyze pitches in the in-course and out-course zones. Figure~\ref{fig:horizontal_zone} visualizes these zones and the changes in decision ratios over the years. The positions differ for left-handed and right-handed hitters, marked with L and R labels in the legend. As shown in Figure~\ref{fig:horizontal_zone}, the strike call ratio increases for pitches within the official strike zone (Out 1, In 1). Conversely, for pitches just outside the strike zone (Out 2, In 2), the robot umpire calls more balls. Notably, in the 2021 and 2022 seasons, left-handed hitters had lower strike-call ratios for in-course pitches compared to right-handed hitters. However, for 2024, there was no statistical difference in strike-call ratios between left- and right-handed hitters, indicating the consistency of robot umpires.


\begin{figure}[t!]
    \centering
    \subfigure[Strike ratios for different pitch types and different years with respect to the pitches located in the out-courses.]{
    \centering
    \includegraphics[width=0.97\linewidth]{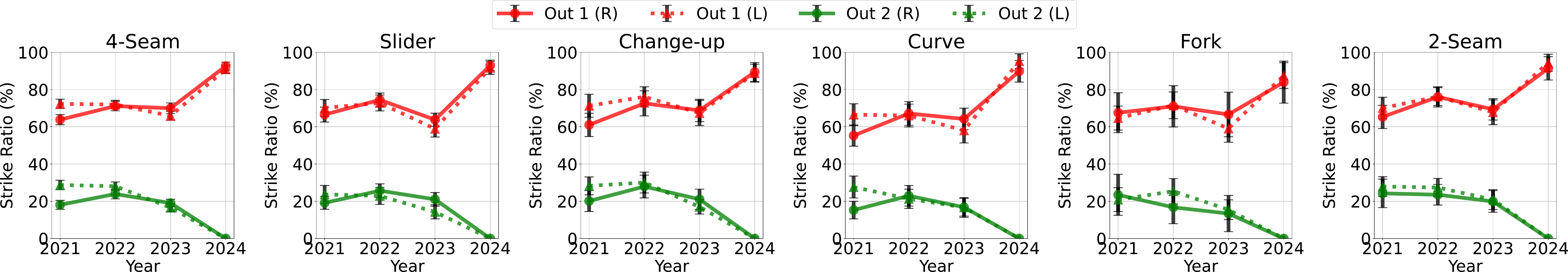}
    }
    \subfigure[Strike ratios for different pitch types and different years with respect to the pitches located in the in-courses.]{
    \centering
    \includegraphics[width=0.97\linewidth]{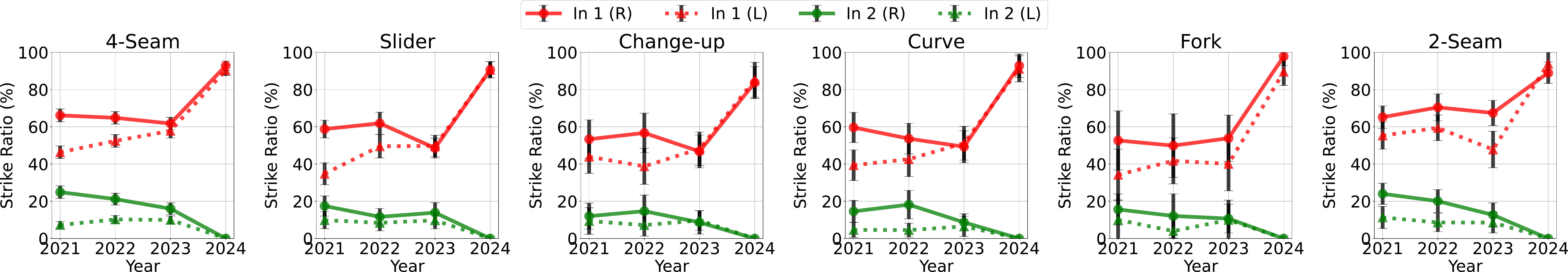}
    }    
    \caption{Trend of strike call ratio concerning the pitch location for different pitch types. The vertical bars in the plots indicate the 95\% credible interval.}
    \label{fig:horizontal_stuff}
\end{figure}

For the horizontal edge zones, we conduct a detailed analysis of pitch classification by human and robot umpires, as illustrated in Figure~\ref{fig:horizontal_stuff}. Figure~\ref{fig:horizontal_stuff}~(a) indicates that, except for forkballs, the strike call ratio for the out-course zone entering the strike zone (Out 1 R and L) significantly increased for all pitch types ($p < 0.05$). Conversely, the strike call ratio for pitches outside the zone (Out 2 R and L) noticeably decreased for all pitch types.

Figure~\ref{fig:horizontal_stuff}~(b) shows that the strike call ratio increased for all pitch types within the strike zone (In 1) regardless of the batter's handedness. Pitches outside the strike zone (In 2) were mostly not judged as strikes. Notably, curveballs and forkballs exhibited the largest increase in strike call ratio. Specifically, the forkball's strike call ratio rose from 53.63\% to 97.72\% for right-handed hitters and from 38.59\% to 91.17\% for left-handed hitters from 2023 to 2024. These results confirm that ABS exhibits narrow (small $\alpha$) but strict (greater $\beta$) strike zones for the horizontal axis.

Using these observations, we offer an answer to \textbf{RQ2} \textit{``Where does the gray zone lie and how do the strike calls differ?"} as follows:
(i) The gray zone was identified at the edges of the strike zone;
(ii) Across all pitch types, there was a decrease in strike calls for pitches outside the rule-book strike zone, coupled with a significant increase in strike calls when pitches entered the rule-book strike zone;
(iii) High balls especially demonstrated a notable increase in strike call ratio when using ABS;
(iv) Pitch types with vertical movement (e.g., curveball, forkball) notably benefited from the use of robot umpires.

\section*{Analysis III: Closer Look at Players Adjustment To ABS}
\label{sec:analysis3}


\begin{figure}[t!]
    \centering
    \subfigure[High-course.]{
    \centering
    \includegraphics[width=0.3\linewidth]{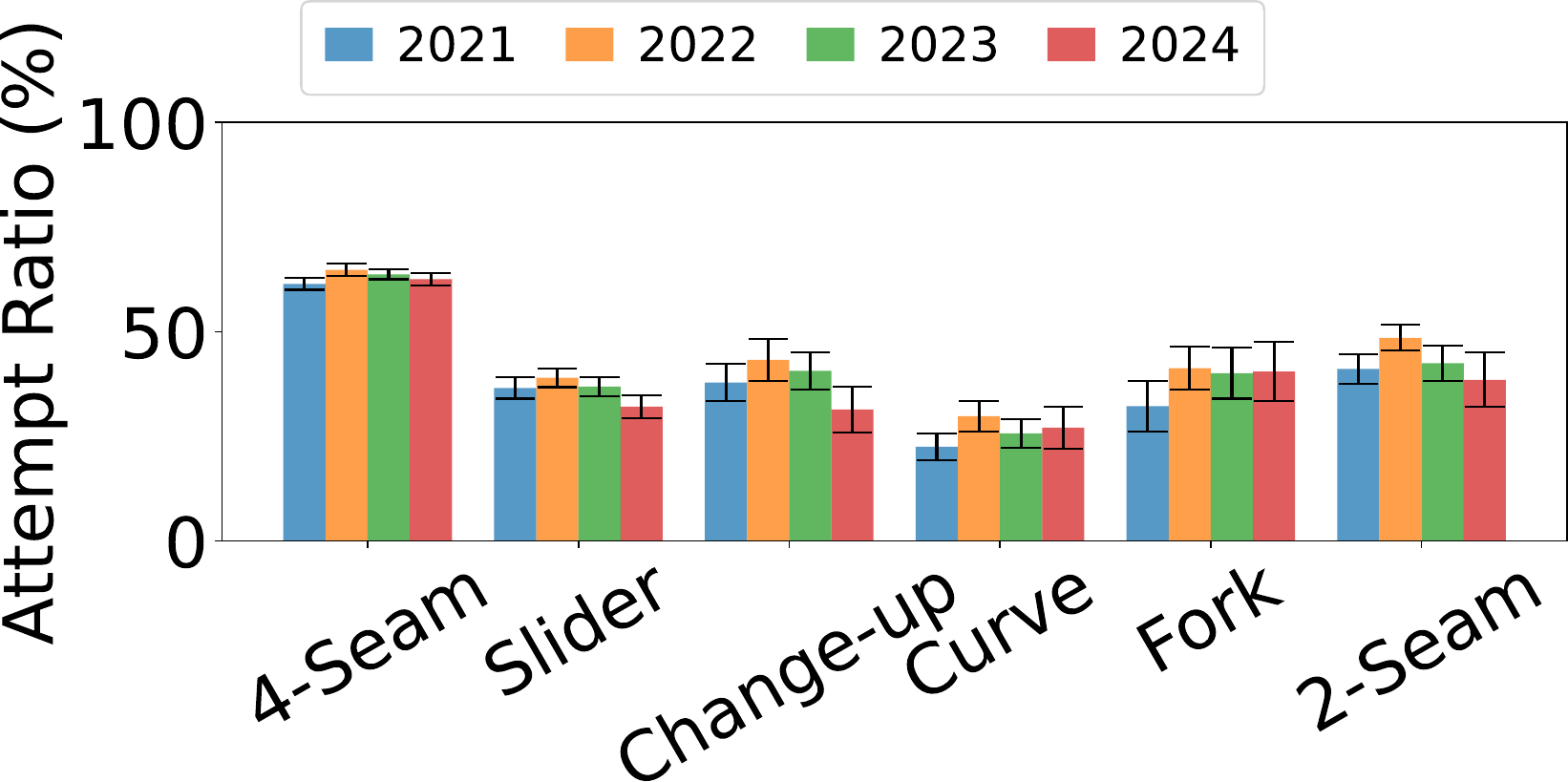}
    }
    \subfigure[Low-course.]{
    \centering
    \includegraphics[width=0.3\linewidth]{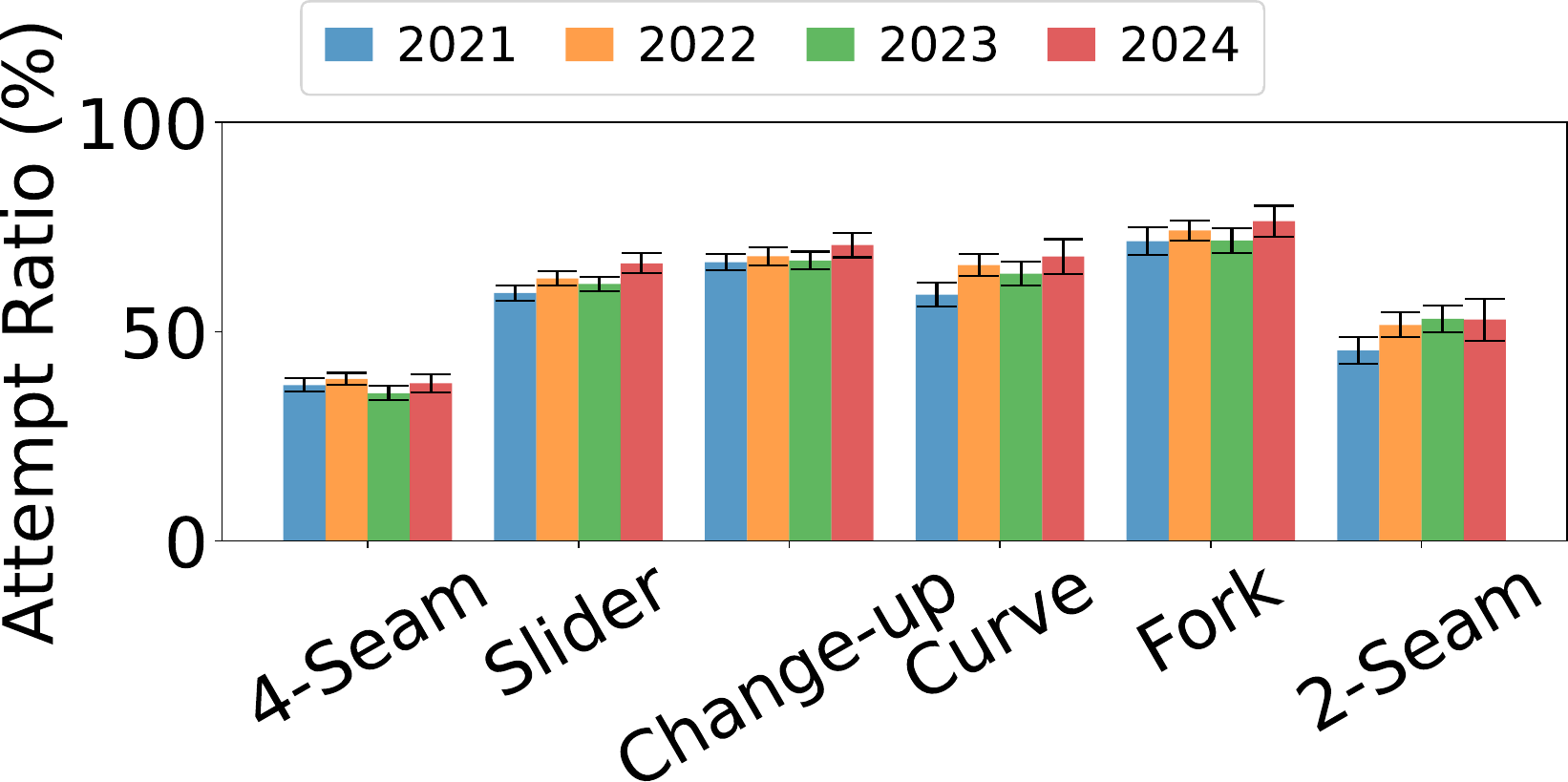}
    }\\
    \subfigure[In-course.]{
    \centering
    \includegraphics[width=0.3\linewidth]{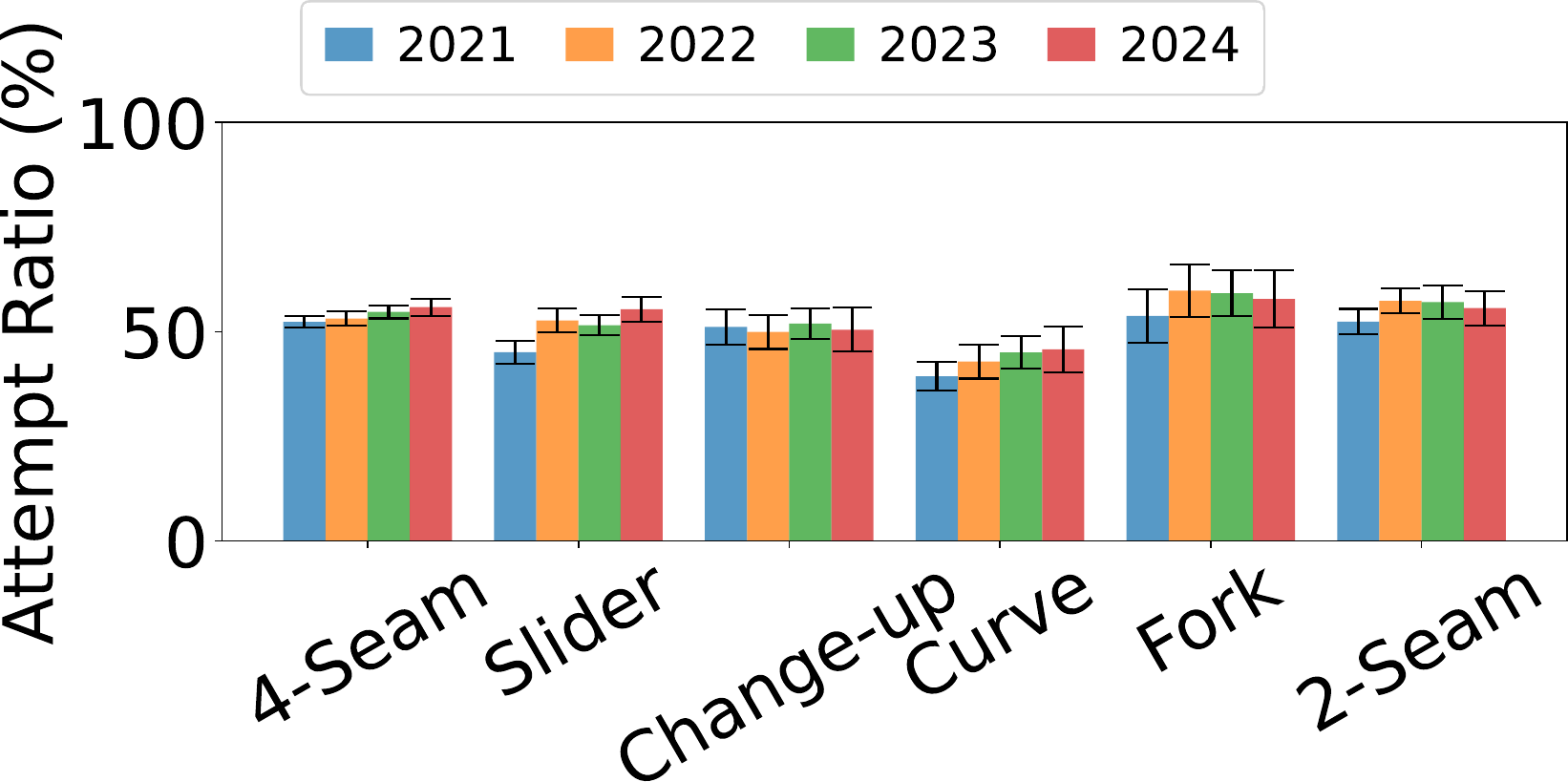}
    }
    \subfigure[Out-course.]{
    \centering
    \includegraphics[width=0.3\linewidth]{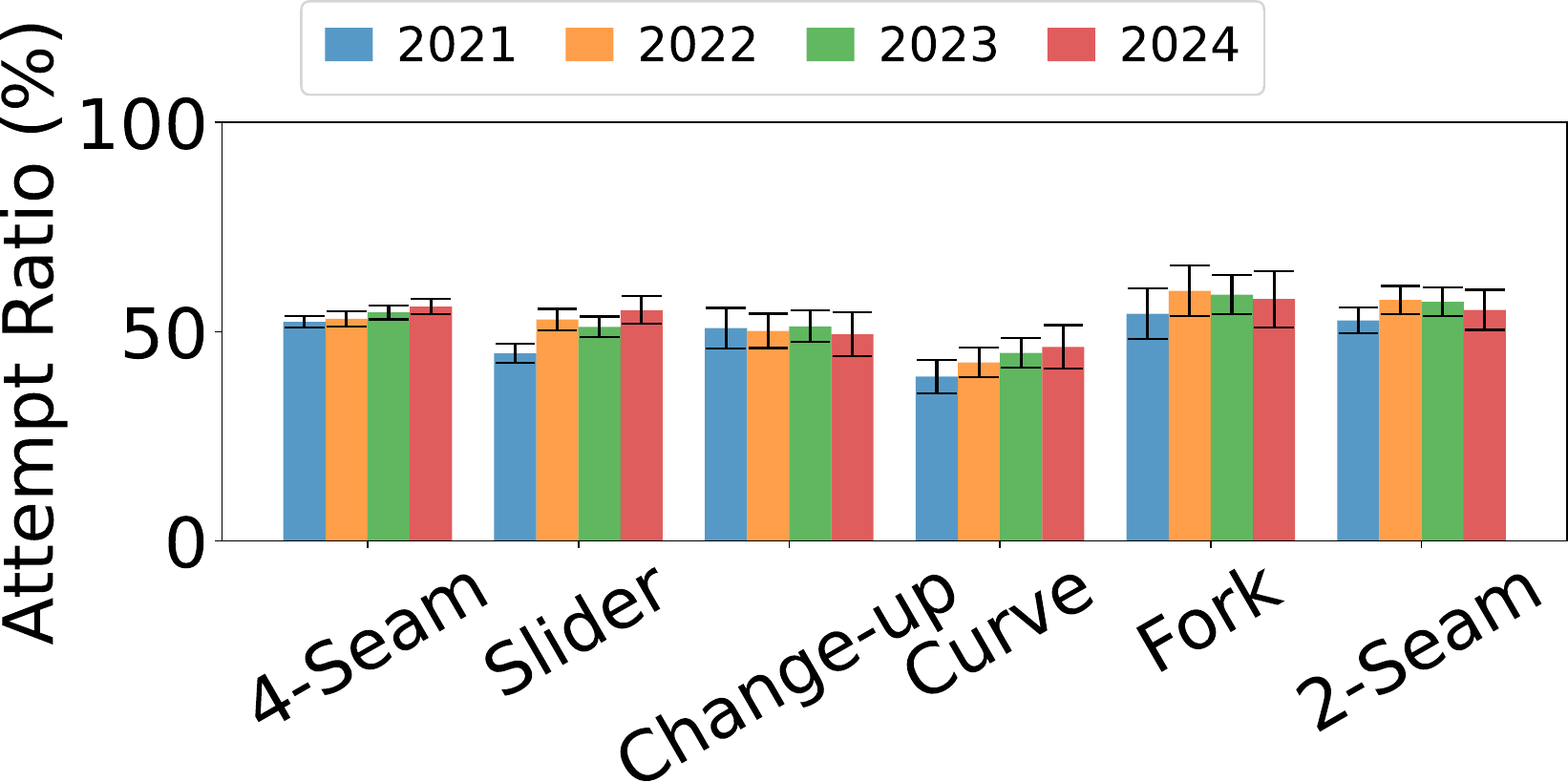}
    }
    \caption{Batters' attempt ratio for different regions and pitch types across years. The vertical bars in the plots denote the 95\% credible interval.}
    \label{fig:hitter_reaction}
\end{figure}

In this section, we examine how batters adjust to the ABS system and how this technology shifts pitchers' strategies. We first quantify batters' responses to pitches in different zones and then analyze the ratio of different pitch types thrown by pitchers in each gray zone discussed earlier.

To evaluate batter responses, we focus on each region inside the rule-book strike zone, considering all swing strikes, hits, fouls, bunts, and bunt fouls as \textit{hit attempts}. Our hypothesis is that if the robot umpire is more lenient with certain pitch trajectories, such as those in the high zone, well-adjusted batters will be more likely to aggressively attempt to hit these pitches.

Figure~\ref{fig:hitter_reaction} presents the hit attempt ratios for different pitch types and regions over various years. Regardless of pitch location, the hit attempt ratios did not show a statistically significant change. Despite the strike call ratio for high-course balls increasing by more than 20\% (cf. Fig.~\ref{fig:vertical_zone}), batters continue to target pitches in the strike zone as they did before ABS was used. This suggests that hitters are not yet properly adjusting to the new system and are still relying on their previous experiences.

\begin{figure}[t!]
    \centering
    \subfigure[High-course.]{
    \centering
    \includegraphics[width=0.3\linewidth]{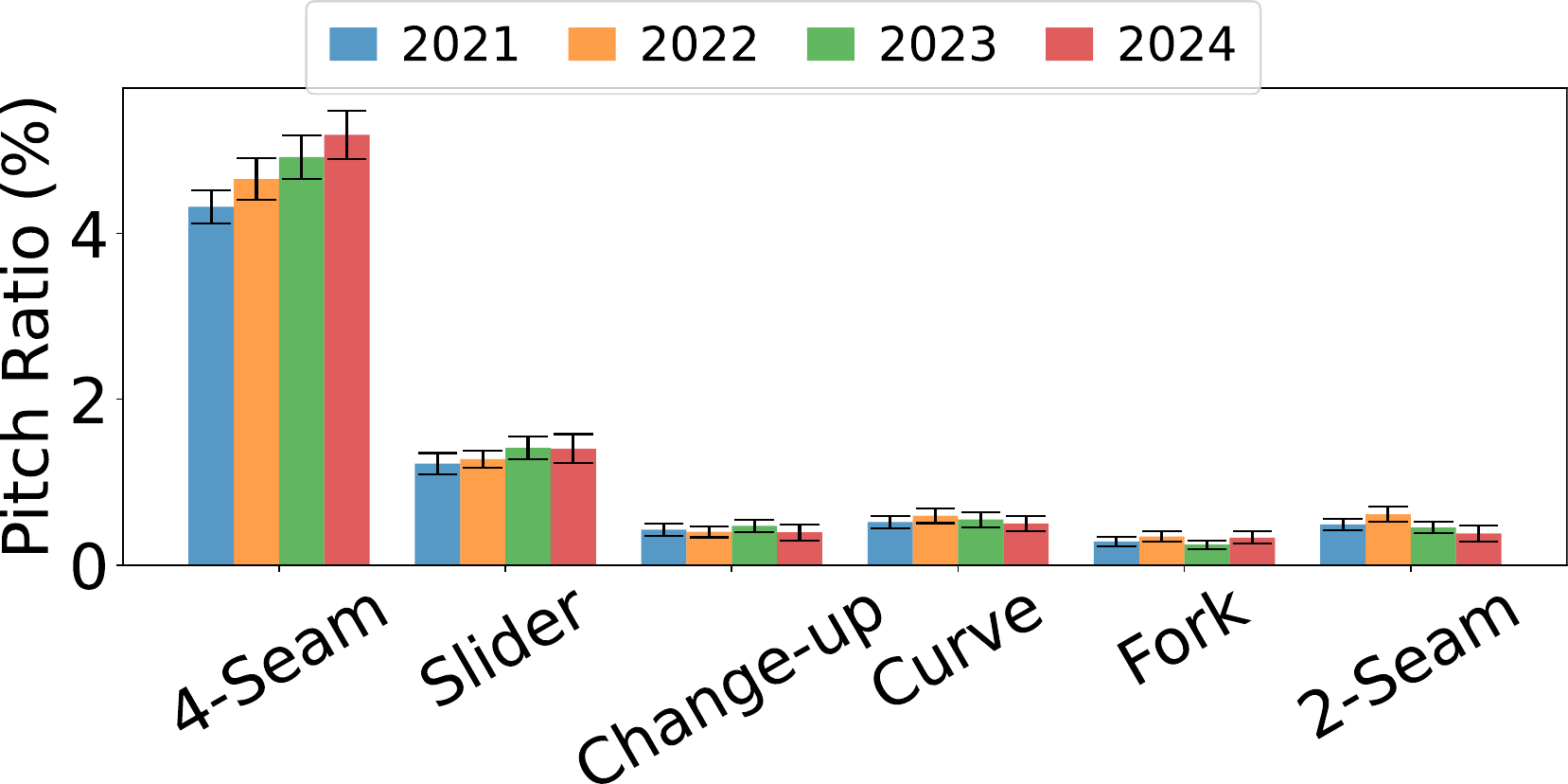}
    }
    \subfigure[Low-course.]{
    \centering
    \includegraphics[width=0.3\linewidth]{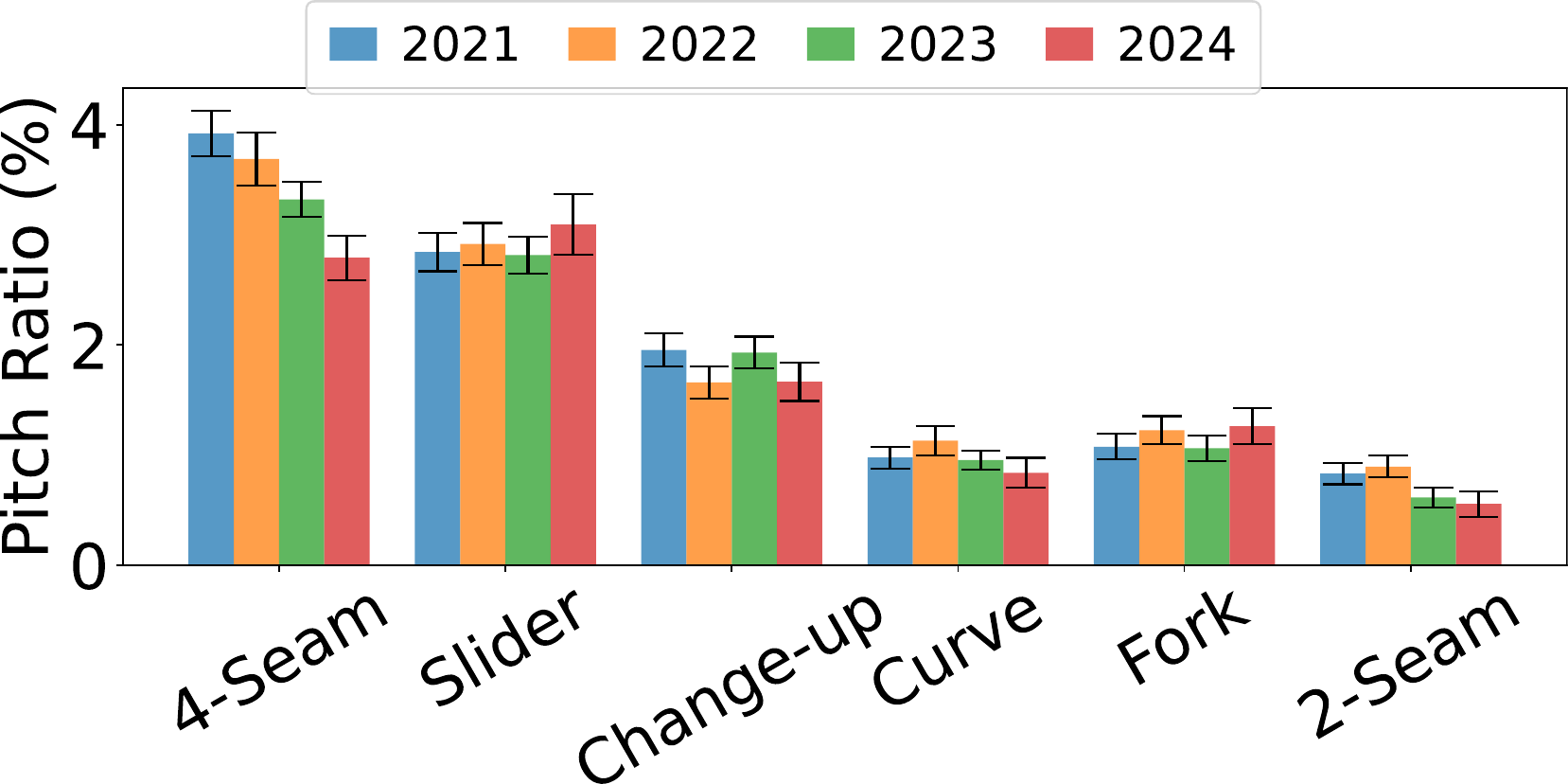}
    }\\
    \subfigure[In-course.]{
    \centering
    \includegraphics[width=0.3\linewidth]{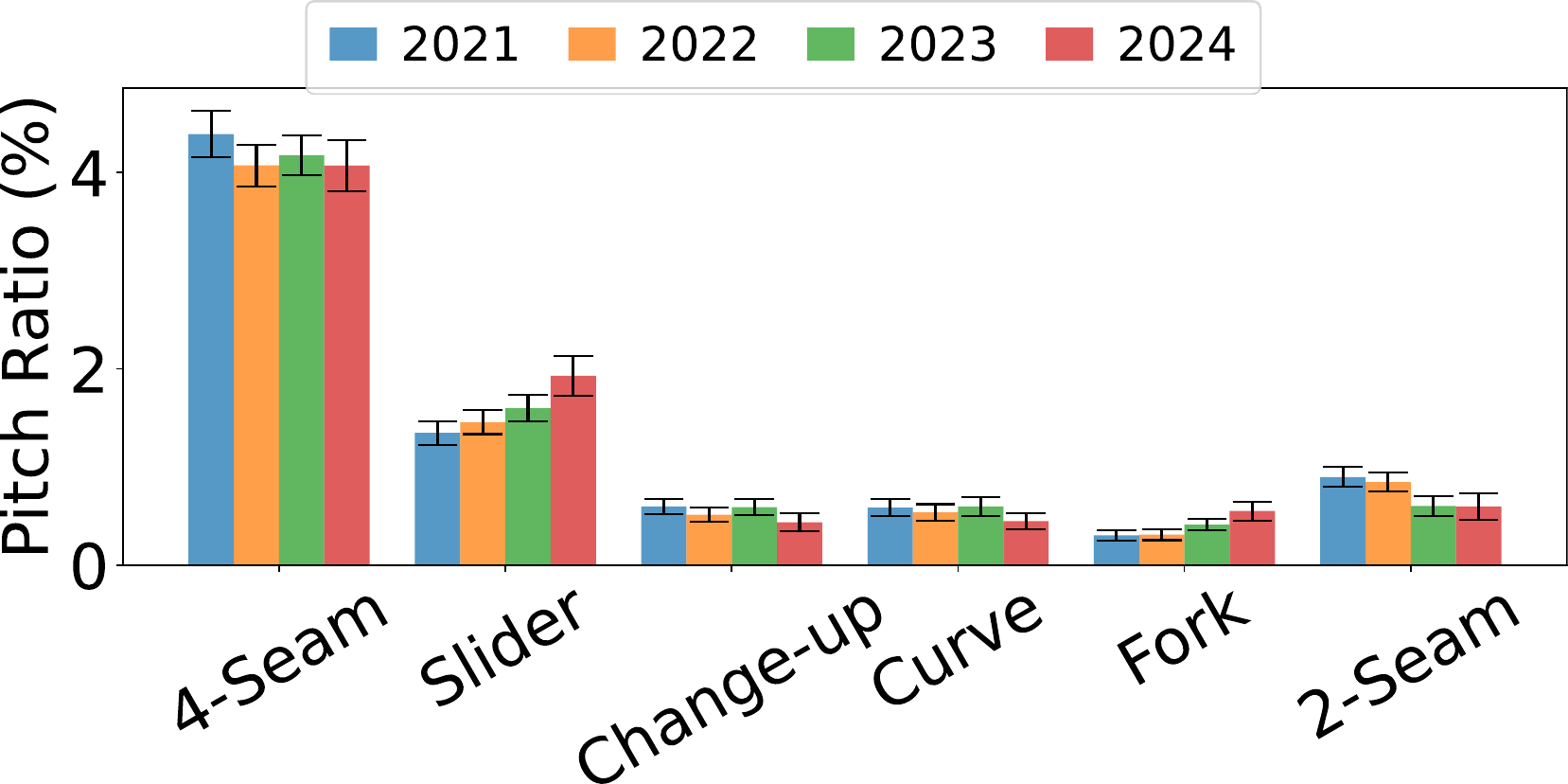}
    }
    \subfigure[Out-course.]{
    \centering
    \includegraphics[width=0.3\linewidth]{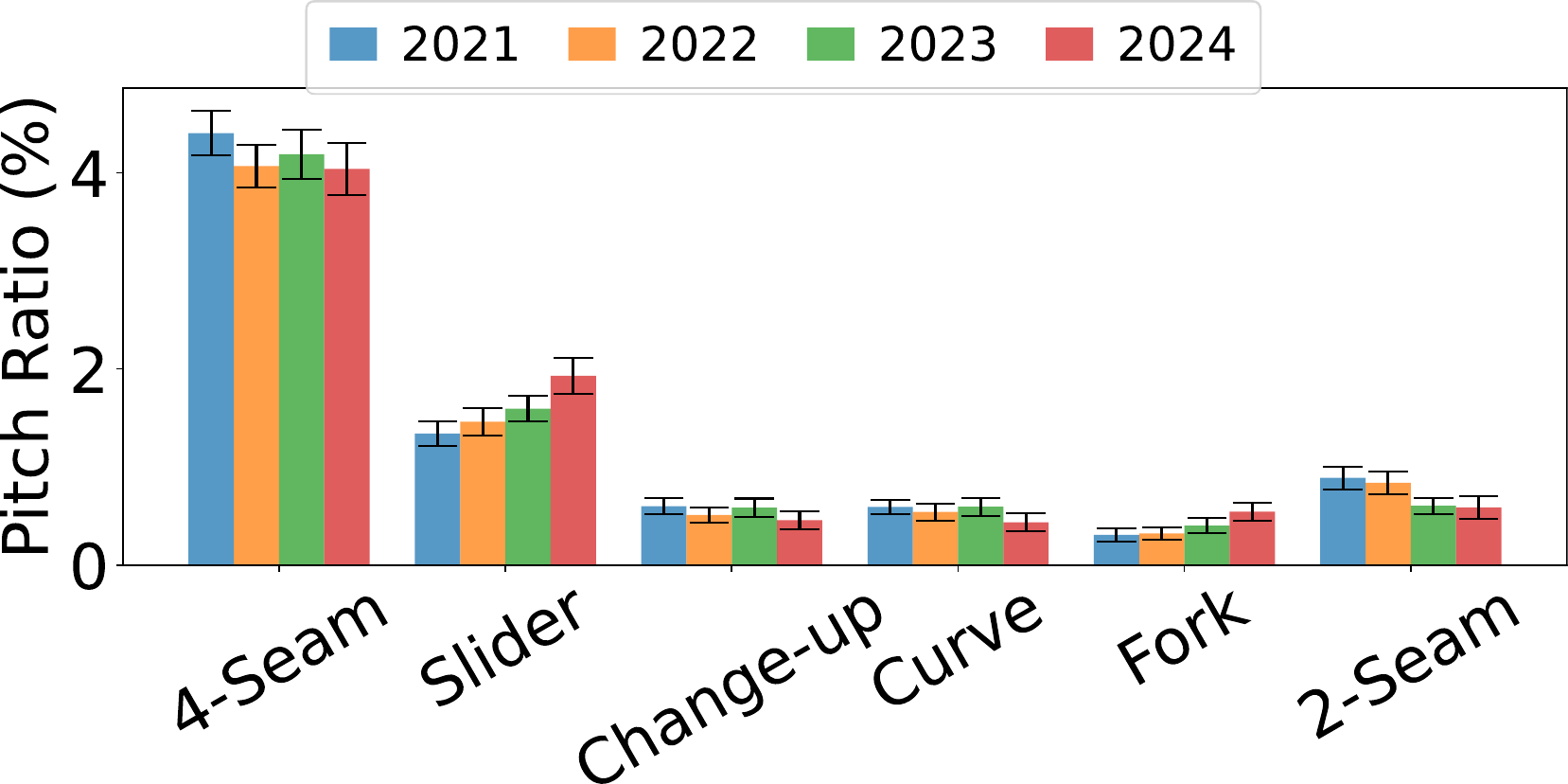}
    }
    \caption{Ratio of pitch for different regions and pitch types across years. The vertical bars in the plots denote the 95\% credible interval.}
    \label{fig:pitch_ratio}
\end{figure}

To understand pitchers' strategic changes, we analyze the frequency of pitches by their arriving zone and types in two balls and two strikes situations, often referred to as decision pitches. Figure~\ref{fig:pitch_ratio} illustrates the pitch ratios of various pitch types thrown in different regions, calculated as the percentage of pitches arriving in each target area.

As Figure~\ref{fig:pitch_ratio}~(a) shows, the 4-seam fastball exhibited an increasing trend in the high regions, while other pitch types did not show significant differences. In Figure\ref{fig:pitch_ratio}~(b), the 4-seam fastball, 2-seam fastball, and curveball exhibited a decreasing trend, while the slider and forkball showed a slight increase. For both lateral zones, as shown in Figures\ref{fig:pitch_ratio}~(c) and~(d), there was no statistically significant change in the frequency for most pitch types. However, sliders, which exhibit prominent transverse movement near the home plate, showed an increased frequency. This suggests that pitchers may be leveraging the slider's movement to exploit the adjusted strike zone boundaries.

The analysis in this section addresses \textbf{RQ3:} \textit{"Are the players adjusting to ABS?"} by examining the batters' hit attempt ratio and the frequency of pitch types. The results reveal that despite the shift in the strike zone, the batters' hit attempt ratio did not change for the gray zones. However, the analysis of pitches thrown in the gray zones indicates that pitchers are relatively exploiting the modified strike zone characteristics.

\section*{Analysis IV: Evaluating The Fairness and Consistency of ABS}
\label{sec:analysis4}

\begin{figure}[t!]
    \centering
    \includegraphics[width=0.85\linewidth]{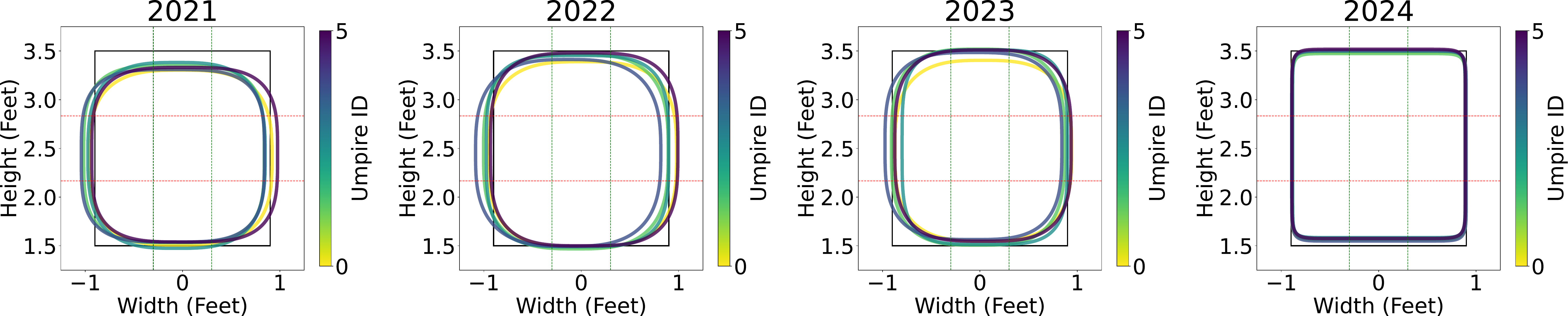}
    \caption{Visualization of the contour lines of strike zones for different umpires in different years. The figure plots the decision boundary (where the probability for calling a pitch as a strike is 0.5) of the fitted model which is identical to the 50\% contour line for each umpire.}
    \label{fig:umpire_zones}
\end{figure}

\begin{figure}[t!]
    \centering
    \includegraphics[width=0.95\linewidth]{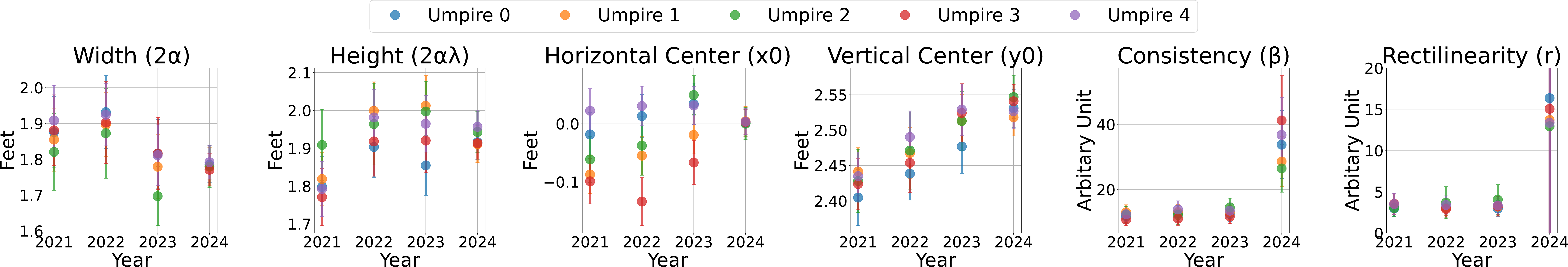}
    \caption{Trend of the fitted model parameter for each umpire across different years from 2021 to 2024. The vertical bars in the plot illustrate the credible intervals in 95\%}
    \label{fig:umpire_params}
\end{figure}

Finally, we investigate the fairness and consistency of the ABS system. ABS aims to reduce human error and bias, which affect the game flow and outcome. We fit each umpire's strike zone and qualitatively compare these zones to demonstrate human umpires' inconsistencies, contrasting these with ABS zones. By examining extensive game data, we assess whether ABS has minimized discrepancies and promoted a uniform and equitable officiating standard.

Figure~\ref{fig:umpire_zones} presents strike zone contour lines for the top five most active home plate umpires from 2021 to 2024. As illustrated, individual umpire strike zones varied across different years, with noticeable discrepancies between umpires in the same season. In contrast, ABS strike zones are consistently dense and uniform, indicating greater consistency and strictness compared to human umpires. This standardization reduces variability and enhances fairness, suggesting that ABS effectively minimizes human error and bias, promoting a more equitable officiating standard in baseball.



We use the logistic function for modeling the strike zone (see Eq.~\ref{eq:pstrike}). Figure~\ref{fig:umpire_params} shows the distribution of strike zone parameters for each umpire's calls, including ABS. The concentrated values in width ($2\alpha$), height ($2\alpha\lambda$), and the center point ($x_{0}$, $y_{0}$) for ABS confirm its consistent and fair strike zone. In contrast, human umpires display statistically significant differences among themselves. The distinct margins in consistency ($\beta$) and rectilinearity ($r$) between human umpires (2021-2023) and ABS (2024) further emphasize the robot umpire's consistency, fairness, and accuracy according to rule-book criteria. The wide credible interval for $r$ arises from the fact that after a certain magnitude, $r$ does not significantly alter the strike zone shape, a characteristic inherent to the p-norm, which requires $r = \infty$ for a perfect rectangle~\cite{biau2015high, alemu2018feedforward}.

\begin{figure}[t!]
    \centering
    \includegraphics[width=0.85\linewidth]{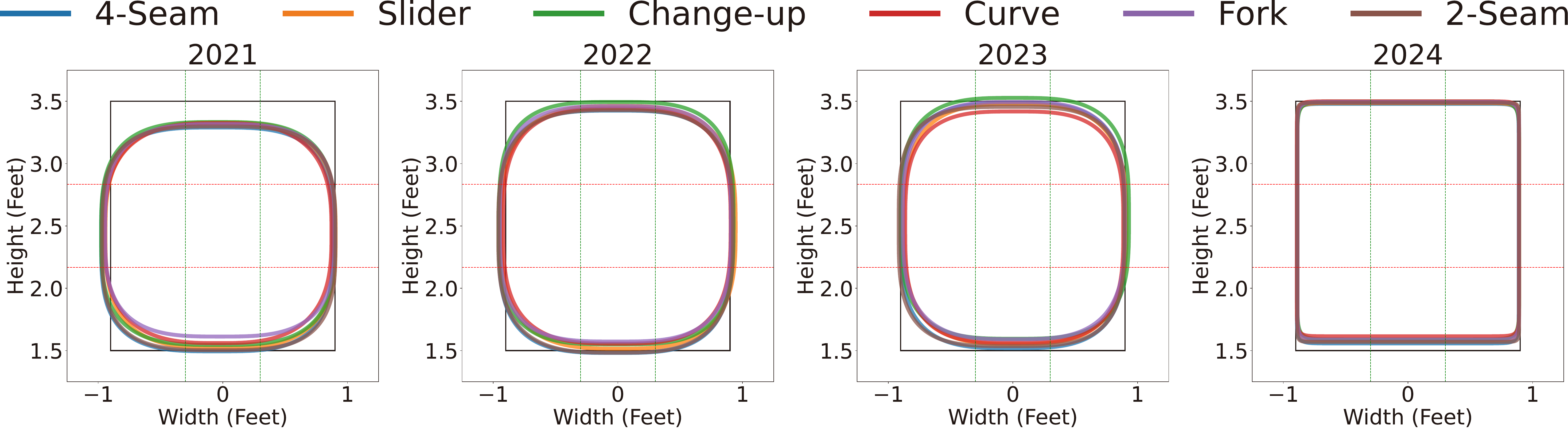}
    \caption{\rev{Visualization of the contour lines of strike zones for different arsenals in different years. The figure plots the decision boundary (where the probability for calling a pitch as a strike is 0.5) of the fitted model, which is identical to the 50\% contour line for each arsenal.}}
    \label{fig:pitchtype_zones}
\end{figure}
\begin{figure}[t!]
    \centering
    \includegraphics[width=0.95\linewidth]{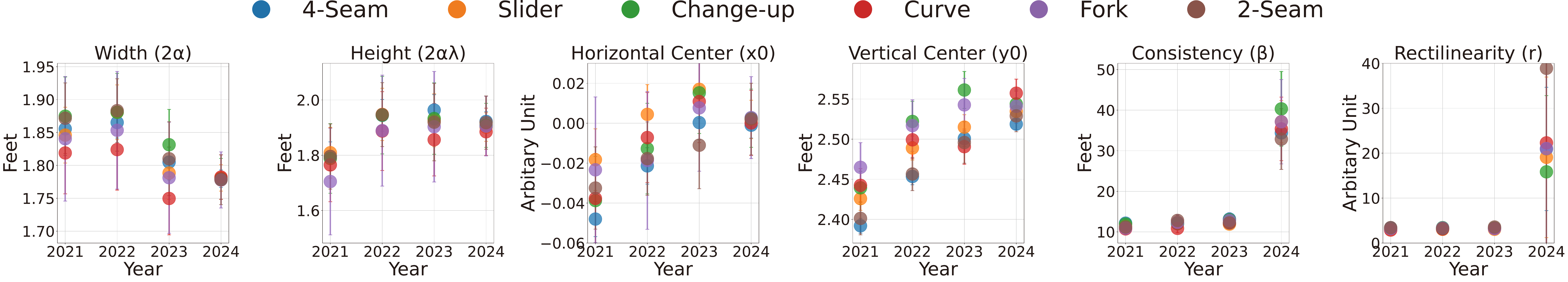}
    \caption{\rev{Trend of the fitted model parameter for each arsenals across different years from 2021 to 2024. The vertical bars in the plot illustrate the credible intervals in 95\%.}}
    \label{fig:pitchtype_params}
\end{figure}

\rev{Furthermore, we investigated whether human umpires and ABS exhibit systematic differences in their treatment of each pitch type by fitting a separate strike-zone model to the calls for each pitch arsenal. Figure~\ref{fig:pitchtype_zones} visualizes the decision boundaries of these per-arsenal models. As the plots show, while ABS deviations across types remain minimal, the human-umpire boundaries display noticeable variability from one pitch type to another. Figure~\ref{fig:pitchtype_params} compares the fitted model parameters across arsenals, demonstrating quantitatively that ABS maintains a tightly clustered zone height ($2\alpha$), width ($2\alpha \lambda$) and center point ($x_{0}, y_{0})$, with significantly high consistency ($\beta$) and rectilinearity ($r$), regardless of pitch type. These findings provide strong evidence that the ABS system delivers more robust and fair verdicts across different pitch arsenals than human umpires.}

Overall, our results highlight ABS's effectiveness in standardizing the strike zone, reducing variability, and minimizing human error and bias, addressing \textbf{RQ4:} \textit{``Is ABS really fair and consistent as expected?''}. This advancement promotes fairness, ensuring consistent standards and enhancing the game's integrity.

\section*{Related work}
\label{sec:relwork}
Pioneering pitch tracking system implemetations such as PITCHf/x~\cite{fast2010heck} and Statcast~\cite{lage2016statcast} have initiated the compilation of extensive pitch tracking and baseball game datasets. The development of these systems and their associated datasets has opened new avenues for statistical analysis in baseball, including pitch sequence analysis~\cite{healey2017using}, pitch movement analysis~\cite{nathan2012determining}, and pitch quality analysis~\cite{hsieh2018graphic, whiteside2016ball}. These advancements have significantly contributed to understanding the intricate dynamics of pitches and player performances, facilitating deeper insights into the sport's strategic and technical aspects.

Furthermore, this rich dataset has not only been utilized to assess the accuracy of umpires' calls~\cite{hunter2018new, fesselmeyer2021impact} but also to investigate additional factors such as player's race~\cite{tainsky2015further, hamrick2015connection}, umpire's age~\cite{flannagan2024psychophysics}, and catcher's expertise~\cite{deshpande2017hierarchical, leblanc2021impact} that may influence game outcomes. Moreover, these studies have contributed to mathematical models of the strike zone, enabling both qualitative and quantitative analyses. For instance, Hunter proposed metrics like Rectangular and Convex hull metrics to assess the strike zone's accuracy and consistency~\cite{hunter2018new}, Deshpande and Wyner developed a hierarchical probabilistic model using Bayesian logistic regression~\cite{deshpande2017hierarchical}, and Flannagan et al. introduced an interpretable parametric logistic function for strike zone modeling~\cite{flannagan2024psychophysics}. In our study, we employ the parametric logistic function-based model to evaluate strike zones. Notably, to the best of our knowledge, previous research has not comprehensively addressed the characteristics of robot umpires (e.g., ABS) and the distinctions between robot and human umpires.

\section*{Discussion}
\label{sec:discussion}
\begin{figure}[t!]
    \centering
    \subfigure[Contour lines of the strike zones for MLB (2024) and KBO (2023, 2024).]{
    \centering
    \includegraphics[width=0.25\linewidth]{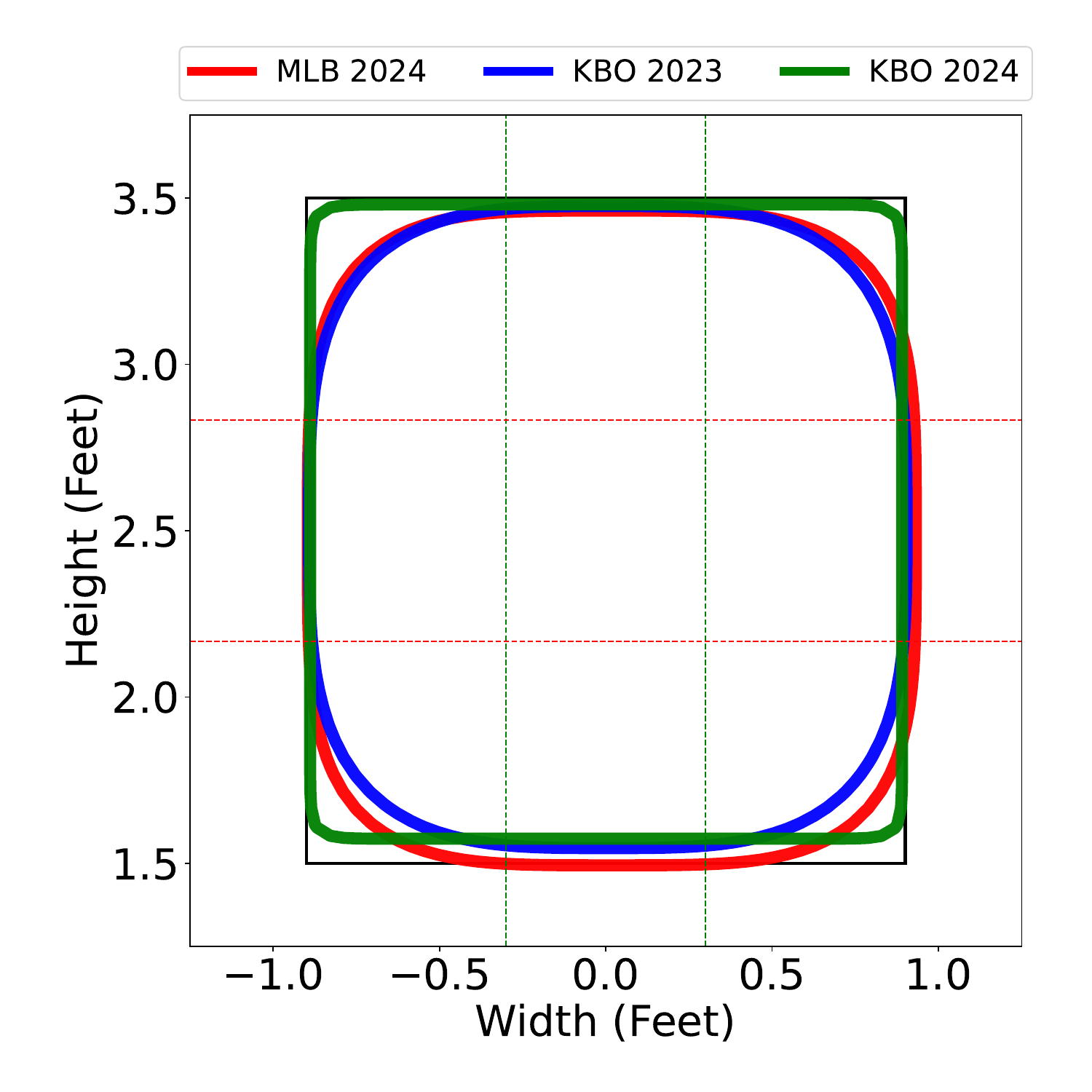}
    }
    \hspace{1ex}
    \subfigure[Difference in calculated strike zone probabilities between MLB (2024) and KBO (2023).]{
    \centering
    \includegraphics[width=0.25\linewidth]{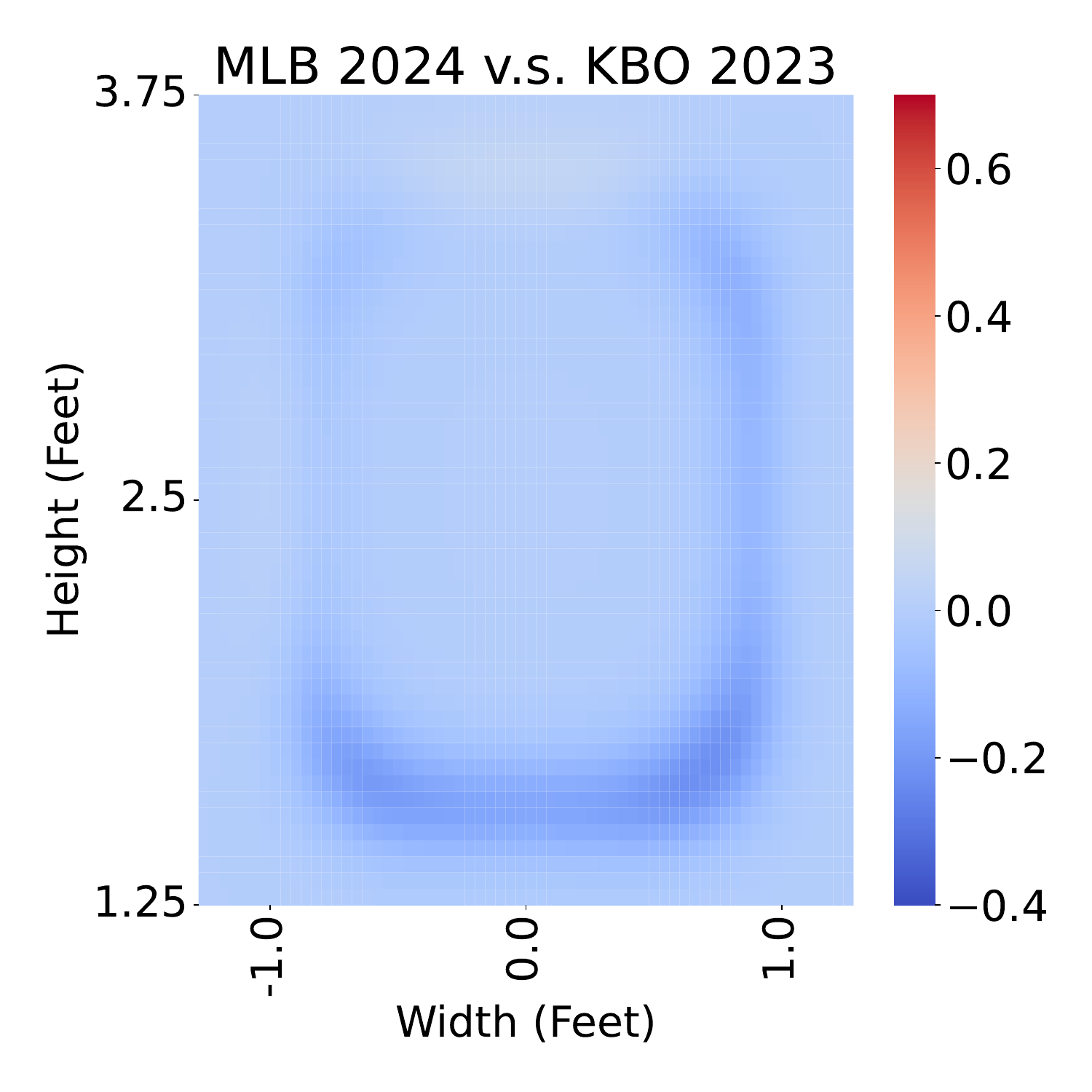}
    }
    \hspace{1ex}
    \subfigure[Difference in calculated strike zone probabilities between MLB (2024) and KBO (2024).]{
    \centering
    \includegraphics[width=0.25\linewidth]{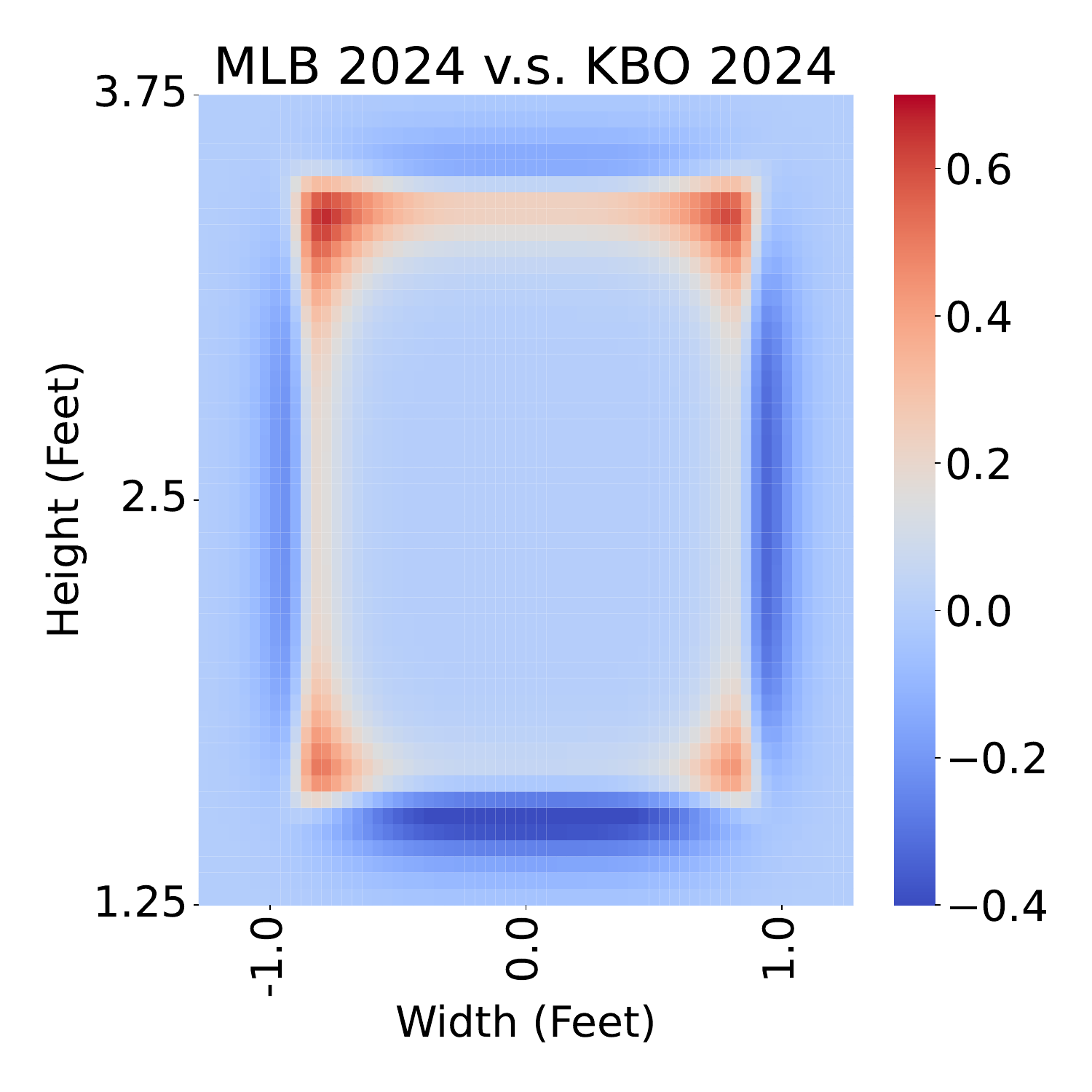}
    }
    \caption{Visualization of the MLB and KBO league strike zone. (a) The figure plots the decision boundary of the fitted model which is identical to the 50\% contour line. (b), (c) The higher/lower value (red/blue) indicates that the KBO umpire (2023, 2024) assigned a high/low probability of strike for the pitches located in the region as strike/ball while the MLB (2024) umpire did not.}
    \label{fig:kbovsmlb}
\end{figure}

\noindent{}$\bullet{}$\textbf{\rev{Potential drawback of ABS.}}
\rev{One limitation of this study is that, by focusing exclusively on differences between human and ABS strike zones, we do not address potential drawbacks of ABS adoption, most notably, the loss of traditional umpiring dynamics and its impact on fan engagement. Removing this human element can diminish the suspense and emotional connection that close calls and umpire reactions provide, potentially detracting from the spectator experience. An interesting direction of future research is to explore how ABS influences fan perceptions and consider strategies (e.g., enhanced broadcast features or mixed‐officiating models) to preserve the game’s dynamic atmosphere.}

\noindent$\bullet{}$\textbf{Per-player level analysis.} 
%
\rev{In this work, we adopted a macroscopic and diachronic view of ABS impact on umpiring decisions. Consequently, in Analysis III where we observed player's adjustment to ABS, we did not perform analysis from a per-player perspective with fine-grained granularity. Note that players can react differently to the ABS system, impacted by factors such as years of experience, batting and pitching style, historical performance metrics, and physical characteristics potentially influencing strategic changes. This would require a detailed persona modeling of each player to capture these individual characteristics. We acknowledge the importance of this deeper, granular analysis and suggest it as a promising direction for future research. Over multiple seasons, for instance, pitchers may shift their pitch-selection toward locations that exploit the ABS consistency, and batters may refine their plate approach in response to a more predictable strike zone. Similarly, coaching staffs and scouts are likely to evolve training regimens and talent evaluation criteria around objective, data‐driven strike‐zone metrics.} Understanding these individualized and systemic responses to ABS will enrich our comprehension of the system’s impact and aid in refining training and adaptation strategies for players.

\noindent$\bullet{}$\textbf{ABS impact in other leagues.} 
While this work focuses on the impact of ABS adoption in the KBO league, assessing its potential effects on other professional baseball leagues, such as Major League Baseball (MLB) and Nippon Professional Baseball (NPB), is an intriguing avenue for future research. Although ABS data are not available for MLB and NPB, we present a preliminary analysis using MLB's 2024 season data (see Tab.~\ref{tab:dataset_config}) and compare it with historical KBO data. Figure~\ref{fig:kbovsmlb} shows the strike zone modeled for the 2024 MLB season using the parametric logistic function approach. Despite differences in leagues and umpires, Figures~\ref{fig:kbovsmlb}~(a) and~(b) reveal strong similarities in human umpire calls between the leagues. Figure~\ref{fig:kbovsmlb}~(c) demonstrates a comparable pattern in strike zone differences between the MLB and KBO ABS data (cf. Fig.~\ref{fig:year_diff}). These findings suggest that the implementation of ABS in MLB could lead to similar player adjustments as observed in the KBO, including changes in strike zone boundaries and consistency.


\noindent$\bullet{}$\textbf{Improving strike zone modeling.} This work employs parametric logistic function-based strike zone modeling~\cite{flannagan2024psychophysics}. However, this method only considers pitch location projected onto a 2D plane from the home plate umpire's perspective, limiting the detailed analysis of the ball's actual movement. Additionally, as noted in \textit{Analysis IV}, the rectilinearity parameter $r$ shows wide credible intervals due to the characteristics of the $p$-norm function~\cite{biau2015high, alemu2018feedforward}. \rev{Moreover, our current approach applies a uniform, league-average zone height rather than adjusting for each batter’s stature, potentially biasing comparisons for very tall or short players.} We suggest that developing a more sophisticated strike zone modeling approach to address these limitations could provide a more comprehensive analysis of baseball pitch data. Such advancements would deepen our understanding of pitch dynamics and the impact of ABS, offering valuable insights for both researchers and practitioners.

\section*{Conclusion}
\label{sec:conclusion}

In this study, we conducted a pioneering analysis of the differences between Automated Ball-Strike (ABS) systems and human umpires \rev{that move beyond prior work, which have been centered on human-umpire biases and player behavior}, offering crucial insights for players and coaching staff. Key findings include:

\begin{itemize} [leftmargin=*]
    \item \textbf{Gray Zones Identified:} We pinpointed ``gray zones'' at the strike zone's edges where discrepancies between human and robot umpires are most pronounced, illustrating ABS's improvements in consistency. \rev{We anticipate that the discrepancies in strike zones identified between human and robot umpires will meaningfully affect players’ experiences and in-game outcomes. To cope with these differences, players will need to adapt and analyze them, while league organizations should both facilitate players’ acclimation to computer-generated judgments and consider revising rules to better align with players’ established standards.}
    \item \textbf{Pitcher and Batter Adjustments:} Pitchers modified their strategies in response to ABS strike zones, while batters' hit attempt ratios remained stable. \rev{These findings suggest that pitchers, who with more active options, are able to capitalize on the consistency of the ABS system, whereas batters do not yet appear to take full advantage of these predictable strike zones. Overall, this implies that batters will benefit from more specialized training, targeted adaptation, and strategic adjustments to effectively respond to ABS officiating.}
    \item \textbf{ABS Fairness and Consistency:} ABS demonstrated greater consistency and fairness compared to human umpires, validating its potential to reduce human error and bias. \rev{We presume that this consistent and fair strike-zone can mitigate the frustration of both players and fans due to erroneous calls.}
\end{itemize}

These findings provide valuable information for teams to better prepare for and adapt to ABS. Understanding the differences in strike calls and the resulting changes in game dynamics will support teams in developing effective strategies under the ABS system. Future research should include long-term, individual-level analyses and expand to other leagues like MLB and NPB to further refine strategies for integrating ABS into baseball. \rev{In addition, more sophisticated analyses of ABS, including biases arising from various environmental conditions and external contextual factors, as well as deeper investigations using three-dimensional trajectory data, would constitute interesting and valuable future research directions.}

\subsection*{Data availability}
Data is provided within the manuscript and can be downloaded through the following URL: \href{https://github.com/eis-lab/where-do-the-robot-umpires-see}{https://github.com/eis-lab/where-do-the-robot-umpires-see}.

\bibliography{reference}

\begin{thebibliography}{10}
\urlstyle{rm}
\expandafter\ifx\csname url\endcsname\relax
  \def\url#1{\texttt{#1}}\fi
\expandafter\ifx\csname urlprefix\endcsname\relax\def\urlprefix{URL }\fi
\expandafter\ifx\csname doiprefix\endcsname\relax\def\doiprefix{DOI: }\fi
\providecommand{\bibinfo}[2]{#2}
\providecommand{\eprint}[2][]{\url{#2}}

\bibitem{sugizaki2023umpire}
\bibinfo{author}{Sugizaki, Y.}, \bibinfo{author}{Nakazato, J.}, \bibinfo{author}{Tsukada, M.} \& \bibinfo{author}{Esaki, H.}
\newblock \bibinfo{title}{Umpire assistance system in baseball game}.
\newblock In \emph{\bibinfo{booktitle}{Proceedings of the 13th International Conference on the Internet of Things}}, \bibinfo{pages}{287--294} (\bibinfo{year}{2023}).

\bibitem{thomas2017computer}
\bibinfo{author}{Thomas, G.}, \bibinfo{author}{Gade, R.}, \bibinfo{author}{Moeslund, T.~B.}, \bibinfo{author}{Carr, P.} \& \bibinfo{author}{Hilton, A.}
\newblock \bibinfo{journal}{\bibinfo{title}{Computer vision for sports: Current applications and research topics}}.
\newblock {\emph{\JournalTitle{Computer Vision and Image Understanding}}} \textbf{\bibinfo{volume}{159}}, \bibinfo{pages}{3--18} (\bibinfo{year}{2017}).

\bibitem{chen2010contour}
\bibinfo{author}{Chen, H.-T.}, \bibinfo{author}{Tsai, W.-J.} \& \bibinfo{author}{Lee, S.-Y.}
\newblock \bibinfo{journal}{\bibinfo{title}{Contour-based strike zone shaping and visualization in broadcast baseball video: providing reference for pitch location positioning and strike/ball judgment}}.
\newblock {\emph{\JournalTitle{Multimedia Tools and Applications}}} \textbf{\bibinfo{volume}{47}}, \bibinfo{pages}{239--255} (\bibinfo{year}{2010}).

\bibitem{lee2020method}
\bibinfo{author}{Lee, H.}, \bibinfo{author}{Kim, J.}, \bibinfo{author}{Kim, J.} \& \bibinfo{author}{Kim, W.-Y.}
\newblock \bibinfo{title}{A method of measuring baseball position at the strike zone}.
\newblock In \emph{\bibinfo{booktitle}{2020 International Conference on Electronics, Information, and Communication (ICEIC)}}, \bibinfo{pages}{1--3} (\bibinfo{organization}{IEEE}, \bibinfo{year}{2020}).

\bibitem{hsieh2024neural}
\bibinfo{author}{Hsieh, J.}
\newblock \bibinfo{journal}{\bibinfo{title}{Neural network-based tracking and 3d reconstruction of baseball pitch trajectories from single-view 2d video}}.
\newblock {\emph{\JournalTitle{arXiv preprint arXiv:2405.16296}}}  (\bibinfo{year}{2024}).

\bibitem{lage2016statcast}
\bibinfo{author}{Lage, M.} \emph{et~al.}
\newblock \bibinfo{journal}{\bibinfo{title}{Statcast dashboard: Exploration of spatiotemporal baseball data}}.
\newblock {\emph{\JournalTitle{IEEE computer graphics and applications}}} \textbf{\bibinfo{volume}{36}}, \bibinfo{pages}{28--37} (\bibinfo{year}{2016}).

\bibitem{gueziec2002tracking}
\bibinfo{author}{Gu{\'e}ziec, A.}
\newblock \bibinfo{journal}{\bibinfo{title}{Tracking pitches for broadcast television}}.
\newblock {\emph{\JournalTitle{Computer}}} \textbf{\bibinfo{volume}{35}}, \bibinfo{pages}{38--43} (\bibinfo{year}{2002}).

\bibitem{tseng2024pitching}
\bibinfo{author}{Tseng, B.-Y.}, \bibinfo{author}{Chiang, H.-T.}, \bibinfo{author}{Chen, J.-L.} \& \bibinfo{author}{Hsieh, H.-C.}
\newblock \bibinfo{title}{Pitching-motion: Pose-based pitch trajectory overlay system}.
\newblock In \emph{\bibinfo{booktitle}{2024 26th International Conference on Advanced Communications Technology (ICACT)}}, \bibinfo{pages}{116--121} (\bibinfo{organization}{IEEE}, \bibinfo{year}{2024}).

\bibitem{russell1997concept}
\bibinfo{author}{Russell, J.}
\newblock \bibinfo{journal}{\bibinfo{title}{The concept of a call in baseball}}.
\newblock {\emph{\JournalTitle{Journal of the Philosophy of Sport}}} \textbf{\bibinfo{volume}{24}}, \bibinfo{pages}{21--37} (\bibinfo{year}{1997}).

\bibitem{kim2014seeing}
\bibinfo{author}{Kim, J.~W.} \& \bibinfo{author}{King, B.~G.}
\newblock \bibinfo{journal}{\bibinfo{title}{Seeing stars: Matthew effects and status bias in major league baseball umpiring}}.
\newblock {\emph{\JournalTitle{Management Science}}} \textbf{\bibinfo{volume}{60}}, \bibinfo{pages}{2619--2644} (\bibinfo{year}{2014}).

\bibitem{buUmpiresMissed}
\bibinfo{author}{Williams, M.~T.}
\newblock \bibinfo{title}{{M}{L}{B} {U}mpires {M}issed 34,294 {B}all-{S}trike {C}alls in 2018. {B}ring on {R}obo-umps? | {B}ostonia --- bu.edu}.
\newblock \bibinfo{howpublished}{\url{https://www.bu.edu/bostonia/2019/mlb-umpires-strike-zone-accuracy/}} (\bibinfo{year}{2019}).

\bibitem{macmahon2008contextual}
\bibinfo{author}{MacMahon, C.} \& \bibinfo{author}{Starkes, J.~L.}
\newblock \bibinfo{journal}{\bibinfo{title}{Contextual influences on baseball ball-strike decisions in umpires, players, and controls}}.
\newblock {\emph{\JournalTitle{Journal of sports sciences}}} \textbf{\bibinfo{volume}{26}}, \bibinfo{pages}{751--760} (\bibinfo{year}{2008}).

\bibitem{sullivan2015astros}
\bibinfo{author}{Sullivan, J.}
\newblock \bibinfo{title}{How the astros wound up with a bigger zone} (\bibinfo{year}{2015}).

\bibitem{deshpande2017hierarchical}
\bibinfo{author}{Deshpande, S.~K.} \& \bibinfo{author}{Wyner, A.}
\newblock \bibinfo{journal}{\bibinfo{title}{A hierarchical bayesian model of pitch framing}}.
\newblock {\emph{\JournalTitle{Journal of Quantitative Analysis in Sports}}} \textbf{\bibinfo{volume}{13}}, \bibinfo{pages}{95--112} (\bibinfo{year}{2017}).

\bibitem{judge2018bayesian}
\bibinfo{author}{Judge, J.}
\newblock \bibinfo{journal}{\bibinfo{title}{Bayesian bagging to generate uncertainty intervals: A catcher framing story}}.
\newblock {\emph{\JournalTitle{Baseball Perspectus}}}  (\bibinfo{year}{2018}).

\bibitem{nbcnewsTechnicalDifficulties}
\bibinfo{author}{NBC}.
\newblock \bibinfo{title}{{T}echnical difficulties are delaying robot umpires in {M}{L}{B} --- nbcnews.com}.
\newblock \bibinfo{howpublished}{\url{https://www.nbcnews.com/news/sports/robot-umpires-mlb-rcna153820}} (\bibinfo{year}{2024}).

\bibitem{espnWhenWill}
\bibinfo{author}{Rogers, J.}
\newblock \bibinfo{title}{{W}hen and how will robot umps arrive in the majors? {L}atest on {M}{L}{B}'s plan --- espn.com}.
\newblock \bibinfo{howpublished}{\url{https://www.espn.com/mlb/story/_/id/40377683/mlb-robot-umpires-automated-balls-strikes-challenge-system-umps-majors}} (\bibinfo{year}{2024}).

\bibitem{ynaExpandsStrike}
\bibinfo{author}{Jee-ho, Y.}
\newblock \bibinfo{title}{{K}{B}{O} expands strike zone for automated system, announces pitch clock rules | {Y}onhap {N}ews {A}gency --- en.yna.co.kr}.
\newblock \bibinfo{howpublished}{\url{https://en.yna.co.kr/view/AEN20240125005400315}} (\bibinfo{year}{2024}).

\bibitem{dongaKoreanBaseball}
\bibinfo{author}{Kang, D.}
\newblock \bibinfo{title}{{K}orean baseball to introduce robot umpires and pitch clocks}.
\newblock \bibinfo{howpublished}{\url{https://www.donga.com/en/article/all/20231020/4497220/1}} (\bibinfo{year}{2023}).

\bibitem{nytimesTripleAGames}
\bibinfo{author}{Stark, J.}
\newblock \bibinfo{title}{{T}riple-{A} games to start fully using automated ball-strike challenge system --- nytimes.com}.
\newblock \bibinfo{howpublished}{\url{https://www.nytimes.com/athletic/5573707/2024/06/18/automated-ball-strike-challenge-system-triple-a/}} (\bibinfo{year}{2024}).

\bibitem{shorturlPreparesRobot}
\bibinfo{author}{Lemire, J.}
\newblock \bibinfo{title}{As mlb prepares for robot umpires, korean baseball organization provides a case study}.
\newblock \bibinfo{howpublished}{\url{https://shorturl.at/SEToJ}} (\bibinfo{year}{2025}).

\bibitem{shorturlUnpacksWhat}
\bibinfo{title}{Sbj unpacks: What mlb can learn from south korea’s robot umps}.
\newblock \bibinfo{howpublished}{\url{https://shorturl.at/8eZiV}} (\bibinfo{year}{2025}).

\bibitem{flannagan2024psychophysics}
\bibinfo{author}{Flannagan, K.~S.}, \bibinfo{author}{Mills, B.~M.} \& \bibinfo{author}{Goldstone, R.~L.}
\newblock \bibinfo{journal}{\bibinfo{title}{The psychophysics of home plate umpire calls}}.
\newblock {\emph{\JournalTitle{Scientific Reports}}} \textbf{\bibinfo{volume}{14}}, \bibinfo{pages}{2735} (\bibinfo{year}{2024}).

\bibitem{whiteside2016ball}
\bibinfo{author}{Whiteside, D.}, \bibinfo{author}{Martini, D.~N.}, \bibinfo{author}{Zernicke, R.~F.} \& \bibinfo{author}{Goulet, G.~C.}
\newblock \bibinfo{journal}{\bibinfo{title}{Ball speed and release consistency predict pitching success in major league baseball}}.
\newblock {\emph{\JournalTitle{The Journal of Strength \& Conditioning Research}}} \textbf{\bibinfo{volume}{30}}, \bibinfo{pages}{1787--1795} (\bibinfo{year}{2016}).

\bibitem{naver}
\bibinfo{author}{Naver}.
\newblock \bibinfo{title}{Baseball : Naver sports --- sports.news.naver.com}.
\newblock \bibinfo{howpublished}{\url{https://sports.news.naver.com/kbaseball/index}} (\bibinfo{year}{2024}).

\bibitem{mlbBaseballSavant}
\bibinfo{author}{Savant, S.}
\newblock \bibinfo{title}{{B}aseball {S}avant: {S}tatcast, {T}rending {M}{L}{B} {P}layers and {V}isualizations --- baseballsavant.mlb.com}.
\newblock \bibinfo{howpublished}{\url{https://baseballsavant.mlb.com/}} (\bibinfo{year}{2024}).

\bibitem{biau2015high}
\bibinfo{author}{Biau, G.} \& \bibinfo{author}{Mason, D.~M.}
\newblock \bibinfo{journal}{\bibinfo{title}{High-dimensional p p-norms}}.
\newblock {\emph{\JournalTitle{Mathematical Statistics and Limit Theorems: Festschrift in Honour of Paul Deheuvels}}} \bibinfo{pages}{21--40} (\bibinfo{year}{2015}).

\bibitem{alemu2018feedforward}
\bibinfo{author}{Alemu, H.~Z.}, \bibinfo{author}{Wu, W.} \& \bibinfo{author}{Zhao, J.}
\newblock \bibinfo{journal}{\bibinfo{title}{Feedforward neural networks with a hidden layer regularization method}}.
\newblock {\emph{\JournalTitle{Symmetry}}} \textbf{\bibinfo{volume}{10}}, \bibinfo{pages}{525} (\bibinfo{year}{2018}).

\bibitem{fast2010heck}
\bibinfo{author}{Fast, M.}
\newblock \bibinfo{journal}{\bibinfo{title}{What the heck is pitchf/x}}.
\newblock {\emph{\JournalTitle{The Hardball Times Annual}}} \textbf{\bibinfo{volume}{2010}}, \bibinfo{pages}{153--158} (\bibinfo{year}{2010}).

\bibitem{healey2017using}
\bibinfo{author}{Healey, G.} \& \bibinfo{author}{Zhao, S.}
\newblock \bibinfo{journal}{\bibinfo{title}{Using pitchf/x to model the dependence of strikeout rate on the predictability of pitch sequences}}.
\newblock {\emph{\JournalTitle{Journal of Sports Analytics}}} \textbf{\bibinfo{volume}{3}}, \bibinfo{pages}{93--101} (\bibinfo{year}{2017}).

\bibitem{nathan2012determining}
\bibinfo{author}{Nathan, A.}
\newblock \bibinfo{journal}{\bibinfo{title}{Determining pitch movement from pitchf/x data}}.
\newblock {\emph{\JournalTitle{Available: baseball.physics.illinois.edu/Movement.pdf}}}  (\bibinfo{year}{2012}).

\bibitem{hsieh2018graphic}
\bibinfo{author}{Hsieh, F.}, \bibinfo{author}{Fujii, K.}, \bibinfo{author}{Roy, T.}, \bibinfo{author}{Hsieh, C.-J.} \& \bibinfo{author}{McCowan, B.}
\newblock \bibinfo{journal}{\bibinfo{title}{Graphic displays of mlb pitching mechanics and its evolutions in pitchf/x data}}.
\newblock {\emph{\JournalTitle{arXiv preprint arXiv:1801.09126}}}  (\bibinfo{year}{2018}).

\bibitem{hunter2018new}
\bibinfo{author}{Hunter, D.~J.}
\newblock \bibinfo{journal}{\bibinfo{title}{New metrics for evaluating home plate umpire consistency and accuracy}}.
\newblock {\emph{\JournalTitle{Journal of Quantitative Analysis in Sports}}} \textbf{\bibinfo{volume}{14}}, \bibinfo{pages}{159--172} (\bibinfo{year}{2018}).

\bibitem{fesselmeyer2021impact}
\bibinfo{author}{Fesselmeyer, E.}
\newblock \bibinfo{journal}{\bibinfo{title}{The impact of temperature on labor quality: Umpire accuracy in major league baseball}}.
\newblock {\emph{\JournalTitle{Southern Economic Journal}}} \textbf{\bibinfo{volume}{88}}, \bibinfo{pages}{545--567} (\bibinfo{year}{2021}).

\bibitem{tainsky2015further}
\bibinfo{author}{Tainsky, S.}, \bibinfo{author}{Mills, B.~M.} \& \bibinfo{author}{Winfree, J.~A.}
\newblock \bibinfo{journal}{\bibinfo{title}{Further examination of potential discrimination among mlb umpires}}.
\newblock {\emph{\JournalTitle{Journal of Sports Economics}}} \textbf{\bibinfo{volume}{16}}, \bibinfo{pages}{353--374} (\bibinfo{year}{2015}).

\bibitem{hamrick2015connection}
\bibinfo{author}{Hamrick, J.} \& \bibinfo{author}{Rasp, J.}
\newblock \bibinfo{journal}{\bibinfo{title}{The connection between race and called strikes and balls}}.
\newblock {\emph{\JournalTitle{Journal of Sports Economics}}} \textbf{\bibinfo{volume}{16}}, \bibinfo{pages}{714--734} (\bibinfo{year}{2015}).

\bibitem{leblanc2021impact}
\bibinfo{author}{LeBlanc, C.}
\newblock \emph{\bibinfo{title}{The Impact of Pitch Level Tracking Data and Catcher Receiving Skills on Strike Calls}}.
\newblock Master's thesis, \bibinfo{school}{The University of Arizona} (\bibinfo{year}{2021}).

\end{thebibliography}
\end{document}